\documentclass{siamltex}
\usepackage{amsmath}
\usepackage{amsfonts}
\usepackage{amssymb}
\usepackage[english]{babel}
\usepackage[dvips]{graphicx}

\def\ve {{\varepsilon }}
\newcommand{\intlim}{{\int\limits}}
\newcommand{\intlimR}{{\int\limits_{-\infty}^\infty}}
\newcommand{\palpha}[0]{\frac{\displaystyle\pi}{\displaystyle 2\alpha}}

\newcommand{\ba}{\begin{eqnarray}}
\newcommand{\be}{\begin{equation}}
\newcommand{\ea}{\end{eqnarray}}
\newcommand{\ee}{\end{equation}}

\def\Im{\mathop{\rm Im~}\nolimits}
\def\Imag{\mathop{\rm Im~}\nolimits}
\def\Re{\mathop{\rm Re~}\nolimits}
\def\Real{\mathop{\rm Re~}\nolimits}
\def\Res{\mathop{\rm Res~}\nolimits}

\def\nn{\nonumber}

\def\re {{\rm e}}
\def\rd {{\rm d}}
\def\ri {{\rm i}}

\def\eps{\epsilon}
\def\w{\omega}

\def\tPhi {{\widetilde \Phi}}

\def\cos{{\rm cos~}}
\def\sin{{\rm sin~}}
\def\0{\phantom{0}}
\def\-{\phantom{-}}
\def\*{\ast}

\def\x {{\bf x}}
\def\u {{\bf u}}

\title{On Budaev and Bogy's
approach to diffraction by the 2D  traction free elastic
wedge\thanks{This work was conducted at London South Bank University
funded by the Industrial Management Committee of the U.K. Nuclear
Licensees under the IMC contract PC/GNSR/5129. Partial funding has
been provided by EPSRC under the grant GR/R13142 and London South
Bank University. Related short reports are {\sc V.~M Babich, V.~A.
Borovikov, L.~Ju. Fradkin, D. Gridin, V. Kamotski and V.~P.
Smyshlyaev}, {\em Diffraction coefficients for surface breaking
cracks}, Proc. IUTAM Symposium held in Manchester, U.K., 16-20 July
2000,  I.~D. Abrahams, P.~A. Martin, and M.~J. Simon, eds., Kluwer
Academic Publishers, Dordrecht, 2000, pp.~209--216; {\sc V.~M
Babich, V.~A. Borovikov, L.~Ju. Fradkin, V. Kamotski and B.~A.
Samokish}, {\em Ultrasonic modelling of tilted surface breaking
cracks}, in J. NDT \& E Int, 37(2) (2003), pp.~105--110. }}

\author{V.~V. Kamotski\thanks{St. Petersburg Department of Steklov
Mathematical Institute, 27 Fontanka, St Petersburg 191023, Russia,
now at Bath Institute of Complex Systems and  Department of
Mathematical Sciences, University of Bath, Bath, BA2 7AY, U.K. (\tt
{v.kamotski@maths.bath.ac.uk}).} \and L.~Ju. Fradkin\thanks{Waves
and Fields Research Group, ECCE, FESBE, London South Bank
University, 103 Borough Rd. London SE1 0AA, (\tt
{fradkil@lsbu.ac.uk}.)} \and  B.~A. Samokish\thanks{Department of
Mathematics and Mechanics, St Petersburg University, 28,
Universitetski Prospekt, Petergof, St Petersburg 198504, Russia.}
\and  V.~A. Borovikov\thanks{Institute of Problems in Mechanics,
Russian Academy of Sciences, 101/1 Prospect Vernadskogo, Moscow
117526, Russia (\tt {root@borovik.msk.ru}).} \and  V.~M.
Babich\thanks{St. Petersburg Department of Steklov Mathematical
Institute, 27 Fontanka, St Petersburg 191023, Russia (\tt
{babich@pdmi.ras.ru}).}}

\begin{document}

\maketitle

\begin{abstract}
Several semianalytical approaches are now available for describing
 diffraction of a plane wave by the 2D (two dimensional)
traction free isotropic elastic wedge. In this paper we follow
Budaev and Bogy who reformulated the original diffraction problem as
a singular integral one.  This comprises two algebraic and two
singular integral equations. Each integral equation involves two
unknowns, a function and a constant. We discuss the underlying
integral operators and develop a new semianalytical scheme for
solving the integral equations.  We investigate the properties of
the obtained solution and argue that it is the solution of the
original diffraction problem.  We describe a comprehensive code
verification and validation programme.
\end{abstract}
\begin{keywords}
{Diffraction, elastic wedge, Sommerfeld transform, singular integral
problem}
\end{keywords}

\begin{AMS}
35J05,35L05,44A15,45Exx,47B35
\end{AMS}

\pagestyle{myheadings} \thispagestyle{plain} \markboth{KAMOTSKI {\it
et al.}}{ DIFFRACTION BY 2D WEDGE}

\section{Introduction}

Sommerfeld (1896) was the first to solve a wedge diffraction problem
-- that of
 diffraction of an {\it electro-magnetic} wave  by
the {\it perfectly conducting semi-infinite screen}. In this famous
paper, he obtained an exact solution  in the form of a superposition
of plane waves propagating in complex directions.  Nowadays this
representation is called the Sommerfeld integral.  Sommerfeld (1901)
also solved the problem of diffraction of an electro-magnetic wave
by the wedge whose angle is a rational multiple of $180^{\rm o}$.
He suggested that since any irrational number can be represented as
the limit of a rational sequence the result  could be extended to
any wedge  angle. The procedure was implemented by Carslaw (1920).

Another analytical method for solving  two-dimensional diffraction
problems was developed by  Smirnoff and Sobolev (1932). They tackled
the  impulse diffraction by an {\it acoustic} wedge under Dirichlet
or
  Neumann  boundary conditions, without  recourse to the frequency domain.  Chapter 12 in Frank and von Mises (1937) which has been written by Sobolev,
   contains a detailed exposition of
the  method and describes its  application to the acoustic wedge.
Friedman (1949$a$, 1949$b$) and Filippov (1959) both applied this
approach to modelling diffraction of an {\it elastic} wave by the
linear semi-infinite crack in an elastic plane. The time-harmonic
version of the problem was treated by Maue (1953) with the
Wiener-Hopf technique.

In the 1950's  Malyuzhinets (1955-1959) solved the problem of  {\it
acoustic} plane wave diffraction by the wedge with impedance
boundary conditions. He sought a solution in the form of the
Sommerfeld integral -- whatever the wedge angle. Then the boundary
conditions have been reformulated in the form of functional,
 difference, equations
 $\label{mal} F(\w\pm\alpha)
= {\rm R}^\pm(\w) F(-\w\pm\alpha), $
 where $F$ is an unknown meromorphic
function and
  ${\rm R}^\pm = (-\sin \w - a^\pm)/(\sin \w-b^\pm),$ with
 $a^\pm$ and $b^\pm$ known
constants.  Malyuzhinets has shown that these equations have an
analytical solution.  Williams (1959) solved the same problem
independently: he took advantage of the  structure of ${\rm R}^\pm$
to simplify the Wiener-Hopf factorisation.

A good exposition of Malyuzhinets' theory as applied to the {\it
acoustic} and {\it electro-magnetic} wedges was made by Osipov and
Norris (1999). The authors have emphasised the features of the
theory which pertain to {\it impedance}, the higher order, boundary
conditions. These lead to a more involved form of ${\rm R}^\pm$
which nevertheless remain rational functions of $\sin \w$ and $\cos
\w$,  allowing for an analytical solution (see also Tuzhilin 1973.)

Several  mathematicians worked on  elastic wedge problems throughout
the 1960's and 70's. An analytical solution was found for the {\it
slippery rigid elastic wedge} -- the wedge with the zero tangential
traction and zero normal displacement on the boundary (e.g. Kostrov
1966).  The three-dimensional smooth {\it elastic wedge with mixed
boundary conditions} also was  treated analytically (e.g. Poruchikov
1986). A comprehensive review of early papers dealing with
diffraction by the elastic wedge was made by Knopoff (1969).

For the {\it traction-free} elastic wedge an analytical solution has
been found only in the degenerate cases of  wedge angles of
 $360^{\rm o}$ (a linear semi-infinite crack -- see Friedman 1949$a$, 1949$b$; Filippov 1959 and Maue 1953) and $180^{\rm o}$ (an elastic half-plane).
  Plane  shear  wave incidence along the bisectrix of
    a quarter plane can also be treated analytically.

There have been many attempts to find an analytical description of
diffraction  by  the elastic wedge in non-degenerate cases. The
problem  became a diffractionist's  analogue of  the famous Fermat's
Last Theorem.  "Diffraction-fermatists" have not been able to obtain
any promising results. It appears that there can be no analytical
solution.  The traction-free elastic wedge problem has to  be
tackled  numerically.

Since the mid  1980's many researchers  followed this route. Several
relied on   potential theory to reduce the problem to  integral
equations. In particular, the problem of  Rayleigh wave diffraction
by the traction-free elastic wedge was attacked this way by Gautesen
(1985 -- 2002$c$)
    and Fujii (1994 and references therein).
 Fujii (1994) studied  the wedge angles between
$36^{\rm o}$ and $180^{\rm o}.$ Gautesen started with the wedge
angles of $90^{\rm o}$  (Gautesen 1985, 1986 and 2002$a$),
$270^{\rm o}$ (Gautesen 2002$b$) and then considered
 the wedge angles restricted to the interval $189^{\rm
o}$ to $327^{\rm o}$ (Gautesen 2001) and to the interval $63^{\rm
o}$ to $180^{\rm o}$ (Gautesen 2002$c$). A detailed exposition of a
version of the   boundary integral equation approach called the
spectral functions method has been given by  Croisille and Lebeau
(1999). The authors consider the challenging problem of diffraction
of a plane acoustic wave by the {\it elastic wedge immersed in
liquid}. The monograph contains both fundamental  developments and
numerical results. The ideas of Croisille and Lebeau have allowed
Kamotski and Lebeau (2006) to reformulate radiation conditions that
are consistent with physical considerations and also allow one to
prove theorems of existence and uniqueness.

In another series of recent papers the problem of diffraction by the
{\it traction-free} elastic wedge was reduced to the singular
integral equations  by representing  the elastodynamic potentials in
the form of the Sommerfeld integrals --  in the spirit of
Malyuzhinets' approach.
  Larsen (1981) appears
to have been the first to attempt to apply this approach to describe
diffraction by the {\it elastic wedge}.  He considered {\it the zero
displacement} boundary condition and reduced the problem to a system
of functional equations  in analytic functions. He suggested that
Chebyshev's polynomials could be used to solve the problem
numerically, but published no further results. Budaev (1995) and
Budaev and Bogy (1995, 1996 and 2002) have taken the
Sommerfeld-Malyuzhinets approach further and reduced the problem to
a system of two functional equations in two analytic functions.
 These  equations are similar to the
ones obtained in the impedance acoustic wedge problem mentioned
above (Malyuzhinets 1955-1959) but are matrix rather than scalar.
Budaev and Bogy (1995-2002) presented numerical results on scatter
of an incident Rayleigh wave, but explanation of their numerical
scheme and some theoretical arguments are vague. Their computed
reflection and transmission coefficients do not always agree with
the corresponding numerical results obtained by other authors.
Nevertheless, as we show in this  paper, Budaev and Bogy's approach
is valid. Our immediate aim is to clarify certain aspects of their
theoretical treatment, reduce the wedge problem to a  singular
integral one and analyze its properties, develop a stable numerical
procedure for its solution and then, assuming the incident field
Rayleigh, evaluate the corresponding diffraction, reflection and
transmission coefficients. We argue that the obtained solution is
that of the original problem.

In \S\S  \ref{statement} and  \ref{reduction} we outline our own
semianalytical recipe for solution of the singular integral problem,
and in \S \ref{newnumsched} we verify and validate the resulting
code. In Appendix \ref{nomenclature} we describe the nomenclature
and in other Appendices, offer the necessary theoretical
considerations, formulas and  numerical options.
\begin{figure}
\begin{picture}(200,150)
\put(80,-100){\includegraphics{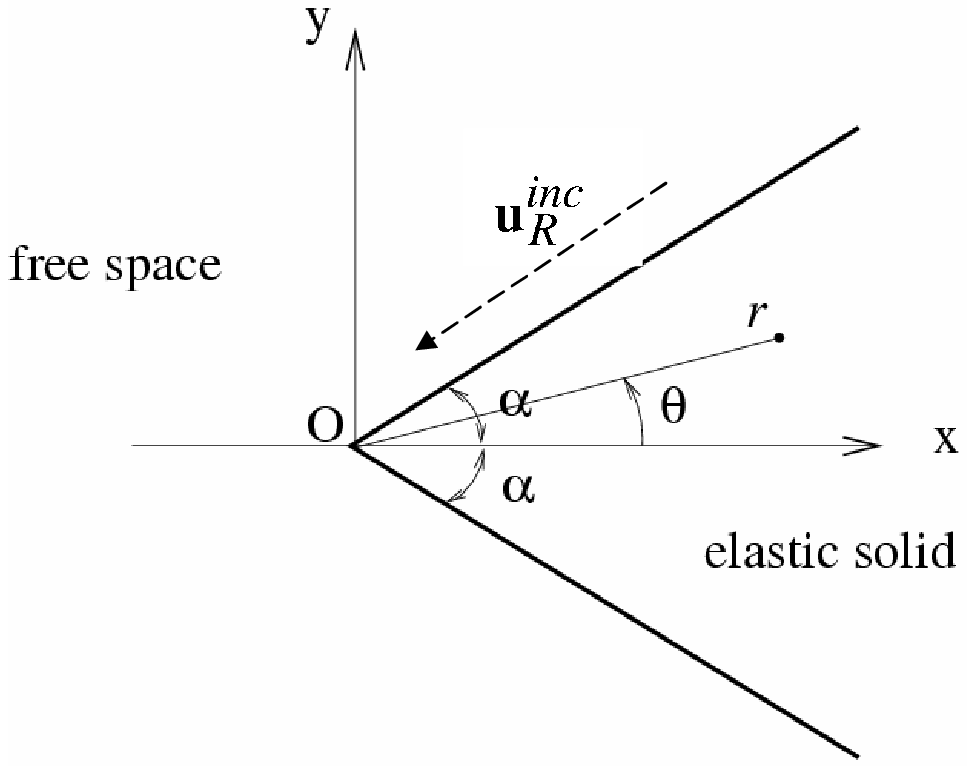}}
\end{picture}
\caption[f]{Geometry of the traction free elastic wedge.}
\label{fig1a}
\end{figure}
\vfill
\newpage
\section{Statement of the problem and the Sommerfeld
amplitudes}\label{statement}
 Let us briefly present the full
statement of the original diffraction problem. We seek the
elastodynamic potentials $\phi_i=\phi_i(kr,\theta)$ that satisfy the
Helmholtz equations in the two-dimensional the wedge of angle
$2\alpha$ with traction free faces, that is, we address the boundary
value problem
\begin{align}
& \Delta\phi_0+\gamma^2k^2\phi_0=0 ,\quad\quad\quad
 \Delta\phi_1+k^2\phi_1=0,\quad &|\theta|<\alpha,\label{helm}\\
 & \Bigg[\frac{2}{r}\frac{\partial^2\phi_0}{\partial
\theta\partial r} +
\frac{1}{r^2}\frac{\partial^2\phi_1}{\partial\theta^2}-\frac{\partial^2\phi_1
}{\partial r^2} +\frac{1}{r}\frac{\partial \phi_1}{\partial r} -
\frac{2}{r^2}\frac{\partial\phi_0}{\partial\theta}\Bigg]=0,\label{tractions1}\
\quad &|\theta|=\alpha,\\
& \frac{1}{\gamma^2}\Bigg[\frac{1}{r^2}
\frac{\partial^2\phi_0}{\partial\theta^2}+\frac{\partial^2
\phi_0}{\partial r^2}+\frac{1}{r}\frac{\partial \phi_0}{\partial
r}\Bigg] -2\Bigg[\frac{\partial^2 \phi_0}{\partial r^2} +
\frac{1}{r}\frac{\partial^2\phi_1}{\partial\theta\partial r} -
\frac{1}{r^2}\frac{\partial\phi_1}{\partial\theta}\Bigg]=0,\quad
\label{tractions2}&|\theta|=\alpha.
 \end{align}
Above and everywhere below, the parameter $k$ is the shear wave
number; $\gamma=c_{\rm S}/c_{\rm P}$ is the ratio of the shear and
compressional speeds $c_{\rm S}$ and $c_{\rm P}$ and the subscript
$i$ takes  values $0$ or $1$.  The geometry of the problem is shown
in Fig.~\ref{fig1a}. Given an incident wave we seek the scattered
potentials satisfying the radiation conditions at infinity
(analogous to the ones in Kamotski and Lebeau 2006, Theorem 4.1) and
bounded elastic energy condition at the wedge tip.

 Note that the
potentials are related to displacement $\u=\u(\x)$ via $\u
=\nabla\phi_0 +\nabla^\perp\phi_1,$ where the nabla operators are
$\nabla = (\partial_x,\partial_y)$, $\nabla^\perp =
(\partial_y,-\partial_x)$.  Note too that in view of this
representation the Helmholtz equations imply
 \be \phi_0(\x)=-(\gamma k)^{-2} \nabla \cdot \u(\x),
 \quad \phi_1(\x)=-k^{-2} \nabla^\perp\cdot\u(\x).
 \label{pot_via_disp}
 \ee
 It follows that  $\phi_i(kr,\theta)$ are uniquely
defined by $\u(\x)$ and since for the corresponding Lam\'{e} problem
in $\u(\x)$,  the  existence and  uniqueness results have been
proven (Kamotski and Lebeau 2006 and Kamotski 2003, Theorem 3.1),
the above diffraction problem as formulated in terms of the
potentials has a unique solution too. Moreover, due to
\eqref{pot_via_disp} all the
 analytical properties of $\u(\x)$ established in Kamotski and Lebeau (2006)
 imply the corresponding properties of
$\phi_i(kr,\theta)$.

\begin{figure}
\begin{picture}(200,100)
\put(150,0){\includegraphics{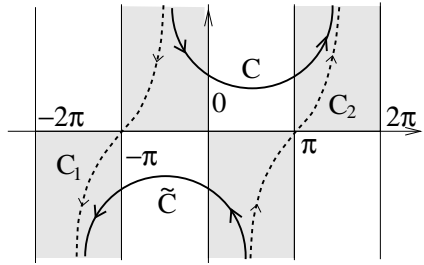}}
\end{picture}
\caption[f]{The Sommerfeld contours in the complex $\w$-plane.}
\label{fig1b}
\end{figure}

The solutions  $\phi_i(kr,\theta)$ of the Helmholtz equations can be
represented in the form of the Sommerfeld integrals
\begin{eqnarray}
\phi_i(kr,\theta) = \int\limits_{C\bigcup{\widetilde C}} \Phi_i
(\w+\theta)\re^{\ri\gamma kr \cos\w}\rd \w, \label{S2int3}
\end{eqnarray}
which can be rewritten as
\begin{eqnarray}
\phi_i (kr,\theta)&=&\intlim_{\rm C}\Big[ \Phi_i (\w+\theta)- \Phi_i
(-\w+\theta)\Big]\re^{\ri\gamma kr \cos \w}\rd \w. \label{S1_int}
\end{eqnarray}
Here   the $\Pi$ shaped contour ${\rm C}$ runs from
$-\pi/2+\ri\infty$ to $3\pi/2+\ri\infty$. The contour ${\widetilde
{\rm C}}$ is  the reflection of ${\rm C}$ with respect to the
origin, and the full integration contour $C\bigcup{\widetilde {\rm
C}}$ is shown in Fig.~\ref{fig1b}. One can justify (\ref{S1_int}) by
applying the Laplace transform in $kr$ to $\phi_i(kr,\theta)$,
changing the Laplace variable $s$ to $\w$, such that  $s = \ri ~
\cos \w$ and exploiting the behavior of $\phi_i(kr,\theta)$ both in
the vicinity of the wedge tip and at infinity (for details see
 Appendix \ref{app:el_pot_repr}). It
follows that the analytic functions $\Phi_i(\w)$ possess a finite
number of singularities  in any vertical strip  and are regular at
the imaginary infinity. Moreover, in Appendix \ref{tip} we show that
the tip behavior of $\phi_i(kr,\theta)$ determines the asymptotic
expansion of $\Phi_i(\w)$ at the imaginary infinity and provides us
with an asymptotic estimate
\begin{equation} \label{reg}
\Phi_i(\w) = O(\re^{-{\Real p}~|\Imag \w |}),~~~\Real p > -1,~~~
|\Imag \w| \to \infty,
\end{equation}
which assures the convergence of integrals \eqref{S1_int}. Note that
these integrals  are invariant under the transformation $\Phi_i (\w
) \rightarrow \Phi_i (\w )+{\rm const}$. In Appendix \ref{tip} we
also show that as $\Imag \w \rightarrow \infty$, the asymptotic
expansion of the preexponential factor in \eqref{S1_int} contains a
constant term. Below, we choose so that the constant terms in their
asymptotic expansions valid as $\Imag \w\rightarrow \infty$ and
$\Imag \w \rightarrow -\infty$ differ only by  sign. We call those
$\Phi_i(\w)$ the Sommerfeld amplitudes.

Introducing the closed contour ${C\bigcup({-\rm
C}_2)\bigcup{\widetilde C}}\bigcup(-{\rm C}_1)$, the Sommerfeld
integrals (\ref{S2int3}) can be evaluated as the contributions of
the poles  and branch points inside that contour plus the sum of
integrals over the steepest descent contours ${\rm C}_1$ and ${\rm
C}_2$.  The former describe all bulk and surface waves arising in
the problem as well as the head waves, and the latter  can be
evaluated using the steepest descent method to provide a description
of the tip diffracted body waves. It follows that all physically
meaningful poles and branch points of $\Phi_i (\w + \theta)$ must be
located   between the contours ${\rm C}_1$ and ${\rm C}_2$ and in
this region  the Sommerfeld amplitudes $\Phi_i (\w )$ should contain
no physically meaningless singularities. Since inside the wedge we
have $|\theta| \le \alpha$, this means that all physically
meaningful singularities
---and only
them---must lie at a finite distance from the horizontal axis
between the contours ${\rm C}_1$ and ${\rm C}_2$ as shifted
horizontally by $-\alpha$
 and $\alpha$ respectively,  that is within the Malyuzhinets region
\begin{equation}
 \{\w:~-\frac\pi2-\alpha-2\tan^{-1} (\re^{-\Imag \w})\le\Real\w\le
 \frac{3\pi}2+\alpha-2\tan^{-1} (\re^{-\Imag \w})\}. \label{A}
\end{equation}
To summarize, the scattered field can be fully described once an
efficient algorithm is produced for calculating the Sommerfeld
amplitudes $\Phi_i(\w)$ in the region \eqref{A}. We proceed with
this task.

\section
{Problem reformulation} \label{reduction} In view of its symmetry
with respect to the polar angle, the original problem naturally
splits into ``symmetric'' and ``antisymmetric'', corresponding
respectively to the symmetric and antisymmetric parts of the
incident wave. These involve functions $\Phi^\pm_i(\w)$, such that
we have
\begin{eqnarray}
\Phi_i^\pm(\w)
=\frac12[\Phi_i(\w)+(-1)^{i+1}\Phi_i(-\w)].\label{sas_deco}
\end{eqnarray}
We follow the approach pioneered in elastodynamics by Budaev (1995)
and first substitute \eqref{S1_int} into the boundary conditions
\eqref{tractions1} and \eqref{tractions2} to obtain the system of
functional equations. We then employ the singular integral
transforms to reformulate the problem as a system of algebraic and
singular integral equations.

\subsection{Functional equations}
 Let us start with the symmetric problem. The boundary conditions \eqref{tractions1}
and \eqref{tractions2} imply
\begin{eqnarray}
& & \intlim_C \gamma^2 a_1(\w) \Big[ \Phi_0^+ (\w+\theta)- \Phi_0^+
(\w-\theta)\Big]\re^{\ri\gamma kr~\cos\w}\rd \w \nonumber\\
& &\qquad\qquad-\intlim_C a_2(\w) \Big[ \Phi_1^+ (\w+\theta)-
\Phi_1^+
(\w-\theta)\Big]\re^{\ri kr\cos \w}\rd\w=0, \label{si11}\\
& &\intlim_C \gamma^2 \Bigg[a_3(\w)+\frac{1}{\gamma^2}\Bigg] \Big[
\Phi_0^+ (\w+\theta)+ \Phi_0^+
(\w-\theta)\Big]\re^{\ri\gamma kr\cos \w}\rd \w \nonumber\\
& &\qquad\qquad-\intlim_C a_1(\w) \Big[ \Phi_1^+ (\w+\theta)+
\Phi_1^+ (\w-\theta)\Big]\re^{\ri kr\cos \w}\rd \w=0, \label{si12}
\end{eqnarray}
where  $ a_1(\w)=\sin 2\w,\ \ a_2(\w)=-\cos2\w$ and $a_3(\w ) = -2
{\rm cos}^2~ \w.$ Introducing in the first terms of both
(\ref{si11}) and (\ref{si12}) the new integration variable $\check
\w$, such that $\cos \check \w=\gamma~\cos \w$, dropping the check
and transforming the contour of integration back to $C$ the
equations acquire the form
 \be \int_C
f(\w)\re^{\ri kr \cos \w}\rd\w = 0, \ee where, due to \eqref{reg},
as $|\Im \w|\rightarrow\infty$, $f(\w) = O(\exp(\Imag \w~[2-\Re p]
))$, with $\Re p>-1$. Using the Malyuzinets theorem (Malyuzhinets
1955) $f$ is an odd trigonometric polynomial of the second order. It
follows that the pair $\Phi^+_i(\w)$ satisfies \eqref{si11} and
\eqref{si12} if and only if it satisfies the functional equations
\begin{eqnarray}
&&t_{11}\Big\{ \Phi_0^+ [g(\w)+\alpha]+ \Phi_0^+
[g(\w)-\alpha]\Big\}+t_{12}\Big\{ \Phi_1^+ (\w+\alpha)+ \Phi_1^+
(\w-\alpha)\Big\}=Q_1^+,\nonumber
\\
\nonumber\\
&&t_{21}\Big\{ \Phi_0^+ [g(\w)+\alpha]- \Phi_0^+
[g(\w)-\alpha]\Big\}+t_{22}\Big\{ \Phi_1^+ (\w+\alpha)- \Phi_1^+
(\w-\alpha)\Big\}=Q_2^+, \nonumber \\ \label{fe1}
\end{eqnarray}
where $t_{11}={\cos 2\w ~\sin\w}/{\sqrt{\gamma^2-{\rm cos}^2~\w}},\,
t_{12}=t_{21}=\sin 2\w,\, t_{22}=-\cos 2\w;$ and we have
\begin{eqnarray}
Q_j^+=c^+_{j1}\sin\w+c^+_{j2}\sin 2\w, \label{defs}
\end{eqnarray}
with $c^+_{jk}, \,j, k=1,2$---unknown constants. The function
$g(\w)=\cos^{-1}(\gamma^{-1}\cos\w)$ relates the incidence shear
angles to compressional reflection angles and its branch cuts are
chosen so that  the deformed contour of integration $\check
C=\{\check\w=g(\w): \w\in C\}$  may be transformed back to $C$
without touching them. They are the segments
\begin{equation}
\label{gcuts} [- \theta_h+\pi n, \theta_h+\pi n],~~\theta_h={\rm
cos}^{-1}~\gamma,\ \ n{\rm ~-~ integer}.
\end{equation}
The branch of $g(\w)$ is chosen so that it has the properties
\begin{eqnarray}
&&g(\frac{\pi}{2}) = \frac{\pi}{2}, \nonumber\\
&&g(-\w) = -g(\w), ~~g(\w+\pi n) = g(\w) + \pi n \nonumber\\
&&g(\w)\simeq\w-\ri \log \gamma+O\Big(\re^{-2|{\rm Im }
\w|}\Big),~~{\rm as~Im}~\w\rightarrow\infty .
\end{eqnarray}

In order to investigate restrictions on $c^+_{ij}$ let us substitute
expansions  \eqref{aspp3} of the Sommerfeld amplitudes into
(\ref{fe1}) and equate the coefficients of the leading asymptotic
terms in the resulting equations. Firstly, we note that in the
symmetric case, (\ref{aspp3}) contains no constant terms and
therefore, there are no $\exp(-2\ri\w)$ terms in the left hand sides
of these equations. This implies that $ c^+_{12}=c^+_{22}=0.$
Secondly, since the tip asymptotics of $\phi^+_i(kr,\theta)$ contain
the terms with the exponent $1$, these sides contain the
$\exp(-\ri\w)$ terms. Equating the coefficients of the
$\exp(-\ri\w)$ terms we obtain \be (\gamma \Phi^+_{0m} +\ri
\Phi^+_{1m} )~\cos\alpha =\frac {\ri}{2}c^+_{11},\quad -(\gamma
\Phi^+_{0m} +\ri \Phi^+_{1m} )~\sin\alpha =\frac{\ri}{2}c^+_{21}.
\ee Hence
  we have
  \be
 c^+_{21}=-c^+_{11}\tan\alpha .
 \ee
Note that if $\gamma \Phi^+_{0m} +\ri\Phi^+_{1m} = 0$, then
$c^+_{21}=-c^+_{11}=0.$

 It follows that
the right hand sides in (\ref{fe1}) {\it might be} --  and as we
show in \S\S \ref{newnumsched} and  \ref{codetesting} {\it are} --
nonzero, so that we have \be Q_1^+ = c_1^+ \sin \w,~~~~~~~~~~~ Q_2^+
= -c_1^+\tan\alpha\, \sin \w , \ee where from now on, for
simplicity, we use  notation $c_1^+=c^+_{11}$.

The antisymmetric problem can be treated similarly, with one minor
modification: For all wedge angles $\alpha$, expansion \eqref{aspp3}
might contain a nonzero constant term $\Phi^-_{00}$ and therefore,
the functional equations might contain a second order term. However,
this term can be eliminated by subtracting $\Phi^-_{00}$ from the
Sommerfeld amplitude $\Phi_0^-(\w)$, redefining it in the process.
The corresponding functional equations are
\begin{eqnarray}
&&t_{21}\Big\{\Phi^-_0[g(\w)+\alpha]+\Phi^-_0[g(\w)-\alpha]\Big\}
+t_{22}\Big\{\Phi^-_1(\w+\alpha)+\Phi^-_1(\w-\alpha)\Big\}=Q_1^-,
\nonumber\\
&&t_{11}\Big\{\Phi^-_0[g(\w)+\alpha]-\Phi^-_0[g(\w)-\alpha]\Big\}
+t_{12}\Big\{\Phi^-_1(\w+\alpha)-\Phi^-_1(\w-\alpha)\Big\}= Q_2^-,\nonumber\\
\label{fe1_asym2}
\end{eqnarray}
with
\begin{eqnarray}
Q_1^- = c_1^- \sin \w,~~~~~~~~~~~ Q_1^- = c_1^-\tan\alpha\, \sin \w
.
\end{eqnarray}
  We note that the above reasoning involves only the
asymptotic terms with $p_m^-=0$    or else  with  $p_m^\pm=1$ and
$N_m^\pm=1$, and therefore applies to  all wedge angles under
consideration. We note too that Budaev and Bogy  have  made several
attempts to establish  restrictions on the constants. They  first
used the arguments of the type outlined above in  Budaev and Bogy
(1998). By excluding from consideration the terms with $p_m^\pm=1$,
it is easy to reach the erroneous conclusion that all constants
$c_{jk}^\pm$ are zero. In the static problems, such exclusion is
justified, because the terms describe body translations. By
contrast, in the dynamic problems, their presence is indicative of
nontrivial phenomena.

We proceed by discussing the singularities of the Sommerfeld
amplitudes. First, we assume that the incident wave is plane or
Rayleigh, so that it  manifests itself in $\Phi^\pm_i(\w)$ in the
form of terms which contain simple poles $\theta_{i\ell}$ in the
strip $|\Re \w|\le \alpha$. The functional equations \eqref{fe1} and
(\ref{fe1_asym2}) can be recast as
\begin{eqnarray}
\left(\begin{array}{l} \Phi_0^\pm(g(\w)+\alpha) \\
\Phi_1^\pm(\w+\alpha)
\end{array}\right)=
&\pm& \left(\begin{array}{ll}
{\rm r}_{11}(\w)&{\rm r}_{12}(\w)\\
{\rm r}_{21}(\w)&{\rm r}_{22}(\w)
\end{array}\right)
\left(\begin{array}{l} \Phi_0^\pm(g(\w)-\alpha)\\
\Phi_1^\pm(\w-\alpha)
\end{array}\right)\nonumber\\
&+& c^\pm_1\frac{\sqrt{\gamma ^2-{\rm cos}^2~\w}}{\Delta(\w)}
\left(\begin{array}{l} e^\pm_1(\w) \\ e^\pm_2(\w)
\end{array}\right),
\label{san_cont}
\end{eqnarray}
where the reflection coefficients for the traction free elastic half
space ${\rm r}_{jk}(\w), \, j,k=1,2$ as well as the Rayleigh
function $\Delta (\w)$ and functions $e_j^\pm(\w)$ are given in
Appendix \ref{nomenclature}. The system (\ref{san_cont}) can be used
to effect the analytical continuation from the strip  $|\Real \w|\le
\alpha$ and thus find all  poles $\theta_{i\ell}$ of the Sommerfeld
amplitudes, with their respective residues, which are located in the
strip $\Real \w\,\epsilon\,I=[\pi/2-\alpha, \,\pi/2+\alpha]$. The
rationale behind the choice of the latter strip is clarified below.
The poles are incidence and reflection angles of the respective
incident, reflected and multiply reflected waves
 and their residues describe the amplitudes of these waves---see Budaev and Bogy (1995, Eqs.~(17) and (18)). The first index in $\theta_{i\ell}$
refers to the  mode of the wave and the second to its place in a
sequence of all incident and (multiply) reflected waves (see Appendix \ref{app:GEpoles}).

Let us now  again follow the above authors and introduce the
decomposition
\begin{eqnarray}
\Phi^\pm_i(\w)=\widehat\Phi^\pm_i (\w)+ \tPhi_i^\pm (\w),~~
\label{*dec}
\end{eqnarray}
where the unknown $\tPhi_i^\pm (\w)$ is regular in the strip
$\Re\w\,\epsilon\,I$, and the known  $\widehat\Phi^\pm_i$ is
\begin{eqnarray}
 \widehat\Phi^\pm_i(\w)=\sum_{\ell} \Res(\Phi^\pm_i;\theta_{i\ell})
 \sigma(\w-\theta_{i\ell}),\quad
 \Re\theta_{i\ell}\,\epsilon\,I.
 \label{stard}
\end{eqnarray}
Above, an otherwise arbitrary function $\sigma(\w)$ should be chosen
to be analytic everywhere inside the strip $\Re \w\,\epsilon\,I$,
except for a simple pole at zero, where it has the residue one. The
weakest restriction we can impose on behavior of $\sigma(\w)$ at the
imaginary infinity is that it grows slower than $\text{exp}(|\Imag
\w|\pi/2\alpha )$. Instead, we impose a stronger restriction---that
it behaves as the amplitudes in  \eqref{reg}. If $\sigma(\w)$
possesses singularities which lie outside the strip $\Re
\w\,\epsilon\,I$ the functions $\widehat\Phi^\pm_i(\w)$ contain
extra poles which describe  waves that are outgoing at physical
reflection angles but have nonphysical amplitudes. This causes no
complication, since the corresponding singular terms in
$\widehat\Phi^\pm_i(\w)$ and $\tPhi^\pm_i(\w)$ mutually cancel.

Next we substitute decomposition \eqref{*dec} into the functional
equations \eqref{fe1} and then (\ref{fe1_asym2}) to obtain the
following inhomogeneous systems of equations for the regular
components of the  Sommerfeld amplitudes,
\begin{eqnarray}
&&\Big\{ \tPhi_0^+ [g(\w)+\alpha]+ \tPhi_0^+
[g(\w)-\alpha]\Big\}+B\Big[ \tPhi_1^+ (\w+\alpha)+ \tPhi_1^+
(\w-\alpha) \Big]=
R_1^++c_1^+ S_1, \nonumber\\
&&A\Big\{ \tPhi_0^+ [g(\w)+\alpha]- \tPhi_0^+
[g(\w)-\alpha]\Big\}+\Big[ \tPhi_1^+ (\w+\alpha)- \tPhi_1^+
(\w-\alpha)\Big]= R_2^++c_1^+\tan\alpha\,S_2,\nonumber\\
\label{fe2a}
\end{eqnarray}
and
\begin{eqnarray}
&&A\Big\{ \tPhi_0^- [g(\w)+\alpha]+ \tPhi_0^-
[g(\w)-\alpha]\Big\}+\Big[ \tPhi_1^- (\w+\alpha)+ \tPhi_1^-
(\w-\alpha) \Big]= R_2^-+c_1^- S_2, \nonumber
\\
&&\Big\{ \tPhi_0^- [g(\w)+\alpha]- \tPhi_0^-
[g(\w)-\alpha]\Big\}+B\Big[ \tPhi_1^- (\w+\alpha)- \tPhi_1^-
(\w-\alpha)\Big]= R_1^--c_1^-\tan\alpha\,S_1,\nonumber\\
\label{fe2b}
\end{eqnarray}
 where we use notations
\begin{eqnarray}
&&A=\frac{t_{21}(\w)}{t_{22}(\w)}=-\tan 2\w,\quad\quad
B=\frac{t_{12}(\w)}{t_{11}(\w)}=\frac{2\cos \w
\sqrt{\gamma^2-{\rm cos}^2\w}}{\cos 2\w},\label{not}\nonumber\\
&&R_1^\pm=-\Big\{\widehat\Phi_0^+[g(\w)\pm\alpha]\pm\widehat\Phi_0^\pm[g(\w)-\alpha]\Big\}-B
\Big[\widehat\Phi_1^\pm(\w+\alpha)\pm\widehat\Phi_1^\pm(\w-\alpha)\Big], \nonumber \\
&&R_2^\pm=-A\Big\{\widehat\Phi_0^\pm[g(\w)+\alpha]\mp\widehat\Phi_0^\pm[g(\w)-\alpha]\Big\}-
\Big[\Phi_1^\pm(\w+\alpha)\mp\widehat\Phi_1^\pm(\w-\alpha)\Big],
\nonumber\\
&&S_1=\frac{\sqrt{\gamma^2-{\rm cos}^2~\w}}{\cos2\w},\quad\quad
S_2=\frac{\sin\w}{\cos2\w}.
\end{eqnarray}

 To
summarize, following Budaev and Bogy, the original problem can be
 reformulated as the following boundary value problem in the theory of analytic
functions: Seek   constants $c_1^\pm$ and  functions
$\tPhi_i^\pm(\w)$, such that
\begin{remunerate}
\item
$\tPhi_i^\pm(\w)$ are analytic for  $\Re \w\,\epsilon\,I$ and
satisfy the asymptotic estimate \eqref{reg},
\item
 the  values that
$\tPhi_i^\pm(\w)$ take on the boundaries of the strip $\Re
\w\,\epsilon\,I$
 are linked by equations \eqref{fe2a} and \eqref{fe2b}
(that is, solve these equations for $\Re\w=\pi/2$).
\end{remunerate}

 The above considerations and the properties of the
Sommerfeld transform which are outlined in Appendix
\ref{app:el_pot_repr} show that such a pair exists if there exists a
solution of the original problem. The uniqueness of $\Phi^\pm_i(\w)$
is a more complicated issue which we address in \S 5.

\subsection{Singular integral equations}
Budaev (1995) has suggested to exploit the fact that for all
functions $F(\w)$ satisfying the first of the above assumptions, the
singular integral transform
 \be \label{hdef}
(HF)(\w)= \frac{1}{2\alpha \ri}
V.P.\intlim_{\pi/2-\ri\infty}^{\pi/2+\ri\infty}
\frac{F(\xi)\rd\xi}{\sin[\frac{\pi}{2\alpha}(\xi-\w)]},~~\Re\w=\frac{\pi}{2}
 \ee
 has the property
 \be
  H:\,F(\w+\alpha)+F(\w-\alpha) \rightarrow
 F(\w+\alpha)-F(\w-\alpha),~~\Re\w=\frac{\pi}{2}. \label{hprop}
 \ee
This means that
 on the vertical line $\Re\w=\pi/2$, the terms in the square brackets in \eqref{fe2a} and \eqref{fe2b}
are related by $H$. The terms in the curly brackets are linked by a
similar explicit transform,
\begin{equation}
{\overline H}F(\w)= \frac{1}{2\alpha \ri}
V.P.\intlim_{\pi/2-\ri\infty}^{\pi/2+\ri\infty} \frac{F(\xi){
g'}(\xi)\rd\xi}{\sin\{\frac{\pi}{2\alpha}[g(\xi)-g(\w)]\}},~~
\Re\w=\frac{\pi}{2}, \label{hdashdef}
\end{equation}
where $g'(\xi)=dg/d\xi.$ This suggests introducing new unknown
functions
\begin{equation}
X^\pm(\w)= \tPhi_0^\pm [g(\w)+\alpha]+ \tPhi_0^\pm [g(\w)-\alpha],
~~Y^\pm(\w)= \tPhi_1^\pm (\w+\alpha)+ \tPhi_1^\pm (\w-\alpha).
\end{equation}
Note that the line $\Re \w=\pi/2$ is of special significance,
because the function $g(\w)$ maps it  onto itself. We can now use
\eqref{hprop}  to transform (\ref{fe2a}) and (\ref{fe2b}) into the
system comprising algebraic equations and singular integral
equations which hold on the vertical line $\Re\w=\frac{\pi}{2}$.
This is the crux of Budaev and Bogy's approach.

Changing to the new  independent real variable $\eta$, such that
$\w=\pi/2+\ri\eta$,  the symmetric problem becomes
\begin{eqnarray}
&&x^+(\eta)+b(\eta)y^+(\eta)=r_1^+(\eta)-
c_1^+\frac{\sqrt{\gamma^2+\sinh^2\eta}}{\cosh2\eta}, \label{ce21}\\
&&a(\eta){\overline {\mathcal H}} x^+(\eta)+{\mathcal
H}y^+(\eta)=r_2^+(\eta)-c_1^+\tan \alpha
\frac{\cosh\eta}{\cosh2\eta}, \label{ce2}
\end{eqnarray}
where we have
\begin{eqnarray}
&&x^\pm(\eta)=X^\pm(\frac{\pi}{2}+\ri
\eta),~~~~y^\pm(\eta)=Y^\pm(\frac{\pi}{2}+\ri \eta).
\end{eqnarray}
Standardizing notations and substituting (\ref{ce21}) into
(\ref{ce2}) the problem transforms to a final {\it singular integral
equation} in two unknowns, a function $y^+(\eta)$  and a constant
$c_1^+$,
\begin{eqnarray}
M^+y^+(\eta) =q_0^+(\eta)+c_1^+q_1^+(\eta), \,\,\eta - {\rm real},
\label{sing_int_eq_+}
\end{eqnarray}
where $M^+={\mathcal H}-a{\overline{\mathcal H}}b$. Using the same
approach the antisymmetric problem transforms to
\begin{eqnarray}
&&a(\eta)x^-(\eta)+y^-(\eta)=r_2^-(\eta)+c_1^-\frac{\cosh
\eta}{\cosh 2 \eta},
\label{al-}\\
&&M^-x^-(\eta) =q_0^-(\eta)+c_1^-q_1^-(\eta), \,\,\eta - {\rm real},
\label{sing_int_eq_-}
\end{eqnarray}
where $M^-={\overline {\mathcal H}}-b{\mathcal H}a.$ The rest of the
nomenclature can be found in Appendix \ref{nomenclature}.

\section{A New Numerical Schedule}
\label{newnumsched} Budaev and Bogy (1995, 1996, 2002) have advanced
various implementations of the  numerical  schedule for computing
$\Phi_i^\pm(\w)$, all of which   involve the following three steps:
\begin{remunerate}
\item evaluating
 $y^+(\eta)$ and $x^-(\eta)$
 on the line $\eta = 0~ (\Re \w =
\pi/2)$ by solving the singular integral equations
(\ref{sing_int_eq_+}) and (\ref{sing_int_eq_-}), and then evaluating
$x^+(\eta)$ and $y^-(\eta)$
 by solving the algebraic equations (\ref{ce21}) and (\ref{al-});
\item evaluating $\tPhi_i^\pm(\w)$  in the strip $\Re \w\,\epsilon\,I$,
using the convolution type transforms (\ref{tPhi}) and (\ref{tPsi})
below, with the  kernels  singular on the boundary of this strip;

\item
continuing the computed Sommerfeld amplitudes  $\Phi_i^\pm(\w)$
analytically to the right of the strip $\Real \w ~\epsilon ~I$ by
using the  functional equations (\ref{san_cont}).  Recastin these
equations to effect the continuation to the left of $\Re
\w\,\eps,\,I$.
\end{remunerate}
We have developed an alternative recipe for carrying out the first
two  steps.

\subsection{Solving  singular
integral equations in two unknowns on the line $\eta=0~(\Re \w =
\pi/2)$}

Let us consider the symmetric case first.  Operator $M^+$ is not
analytically invertible, but Budaev and Bogy (1995) suggest that
Eq.~(\ref{sing_int_eq_+}) can be rewritten as
\begin{equation}
({\mathcal H}d+K)y^+(\eta)=q_0^+(\eta)+c_1^+q_1^+(\eta),
\label{H'dy}
\end{equation}
where ${\mathcal H}$ is the  singular operator introduced above,
analytically invertible in the space of bounded functions; $K$ is a
regular operator; and $d(t)$ is
 an exponentially   decreasing function.
Importantly, ${\mathcal H}$ has the property
\begin{eqnarray}
\intlimR {\mathcal H}f(\eta)d\eta = 0,\label{zer_int}
\end{eqnarray}
and therefore, its range consists of all $L^2(\mathbb R)$ functions,
with the zero integral, where  $L^2(\mathbb R)$ is the space of all
integrable functions of real variable.
 Budaev and Bogy
(1996) state that they regularize (\ref{H'dy})
 by applying  ${\mathcal H}^{-1}$ to both its sides.
They carry out numerical evaluation of the resulting singular
integral equation by using (\ref{zer_int})   as  a constraint and
calculate $c_1$ and $y(\eta)$ both at once.  By contrast, below we
argue that the right hand side of (\ref{H'dy}) belongs to the domain
of ${\mathcal H}^{-1}$  for only one value of $c_1$,   carry out the
regularization by finding this value and thus arriving at a singular
integral  equation in one unknown, $y(\eta)$.
 At present, our schedule
works only for $\alpha < \pi/2.$

We start by observing that Eq.~(\ref{H'dy}) is solvable only if its
right hand side belongs to the range of  ${\mathcal H}d+K$. We
cannot describe this range explicitly.  However, it is clear that
$(-Ky^++q_0^++c_1^+q_1^+)(\eta)$ should be in the range of
${\mathcal H}$. It follows that we must have
\begin{eqnarray}
\intlimR
\left[(Ky^+)(\eta)-q_0^+(\eta)-c_1^+q_1^+(\eta)\right]\,d\eta = 0.
\end{eqnarray}
{\bf All our numerical experiments confirm that neither
$q_0^+(\eta)$ nor $q_1^+(\eta)$ are in the range of ${\mathcal
H}d+K$}---by producing the nonzero ``solution defects''
$\lambda_0^+$ and $\lambda_1^+$  defined by (\ref{lambdas}) below.
Therefore,  Eq.~\eqref{H'dy} is solvable only if  the right hand
side of \eqref{H'dy} is in the range. This gives the following
relationship between $c_1^+$ and $y^+(\eta)$:
\begin{equation}\label{c_1+}
c_1^+={\displaystyle
\intlimR\left[(Ky^+)(\eta)-q_0^+(\eta)\right]\,d\eta}
\Big[{\displaystyle \intlimR q_1^+(\eta)\,d\eta}\Big]^{-1}.
\end{equation} By substituting \eqref{c_1+} into \eqref{H'dy}, $c_1^+$
is eliminated and we obtain
\begin{equation}\label{H'dyP}
({\mathcal H}d+P_{q_1^+}K)y^+(\eta)=P_{q_1^+}q_0^+,
\end{equation}
where an unbounded projector
\begin{equation} (P_{q_1^+}u)(\eta)=u(\eta)-q_1^+(\eta){\displaystyle
\intlimR u(t)\,dt}\Big[{\displaystyle\intlimR
q_1^+(t)\,dt}\Big]^{-1}
\end{equation}
maps any function in $L^2(\mathbb R)$ with a finite integral into
the range of ${\mathcal H}$ and has the property
\begin{eqnarray}\label{Pf0_prop1}
P_{q}u(\eta)= \left\{
\begin{array}{ll}
u(\eta),&\text{for all {\it u}}(\eta), \text{such that }\intlimR u(t)\,dt=0,     \\
0, \,\,&\text{for} \,\,\,u(\eta)=q(\eta).
\end{array} \right.
\label{Pf0_prop2}
\end{eqnarray}
We refer  to the function $q(\eta)$ as the projector kernel.

Note that in \eqref{H'dyP}, the integrals of both
$P_{q_1^+}Ky^+(\eta)$ and $P_{q_1^+} q_0^+(\eta)$ are zero, and
therefore, the inverse operator ${\mathcal H}^{-1}$ can now be
safely applied to both sides. Introducing on top of that a new
unknown function $\widetilde y^+(\eta)=d^{1/2}(\eta)y^+(\eta)$ the
final regularized integral equation is
\begin{equation}\label{int_eq_main_+}
\widetilde y^+(\eta)+\widetilde L^+\widetilde y^+(\eta)=\widetilde
q^+(\eta),
\end{equation}
where  $\widetilde L^+={\displaystyle d^{-1/2}}{\mathcal H}^{-1}
P_{q_1^+}K{\displaystyle d^{-1/2}}$ is  an operator with a smooth
kernel and the right-hand side $\widetilde q^+(\eta)={\displaystyle
d^{-1/2}}(\eta) {\mathcal H}^{-1}{P_{q_1^+} q_0^+}(\eta)$. The
equation involves one unknown,  ${\widetilde y}^+(\eta)$, and can be
solved using  a standard quadrature method (see e.g.~Atkinson 1977).
Note that normalizing the original unknown by $d^{1/2}(\eta)$ rather
than $d(\eta)$ leads to a new operator with a bounded kernel ({\it
cf.} Budaev and Bogy 1995). The normalization achieves
symmetrization of the kernel, so that  whether
 $\eta\rightarrow \infty$ or $t \rightarrow \infty$, it exhibits the same
singular behavior.

Eqs.~\eqref{c_1+} and \eqref{H'dyP} imply \eqref{H'dy}.  This means
that the combination of $c_1^+$ and a solution of \eqref{H'dyP}
gives us the solution of \eqref{H'dy}. However, in our code instead
of solving (\ref{int_eq_main_+}) we implement a slightly different
approach: Since $q_1^+(\eta)$ is rather complex, instead of
$P_{q_1^+}$ we employ the projector $P_{q_2}$, with the kernel
\begin{equation}
  q_2(\eta)=\frac{1}{2\alpha}\frac{1}{\cosh\frac{\displaystyle \pi}
  {\displaystyle 2\alpha}\eta}. \end{equation}
Numerical experiments have shown that this  kernel leads to a
  stable evaluation scheme.
We  then regularize and solve two equations
\begin{eqnarray}\label{H'dyPractice1}
({\mathcal H}d+P_{q_2}K)y_i^+(\eta)=P_{q_2}q_i^+(\eta),\,\,i=0,1
\label{H'dyPractice2}
\end{eqnarray}
(see Appendix \ref{inteqns}). The ``solution defects''
\begin{eqnarray}
\lambda_i^+=\intlimR\bigl[(Ky_i^+)(t)-q_i^+(t)\bigr]dt, \,\,i=0,1
\label{lambdas}
\end{eqnarray}
turn out to be nonzero indicating that neither $q_0^+(\eta)$ nor
$q_1^+(\eta)$ are in the range of ${\mathcal H}d+K$.  It follows
that the   solution $(y^+(\eta), c_1^+)$ of
Eq.~(\ref{sing_int_eq_+}) can be  obtained using
\begin{eqnarray}
c_1^+=-\frac{\lambda_0}{\lambda_1},\,\,\,
y^+(\eta)=y_0^+(\eta)+c_1^+y_1^+(\eta).
\end{eqnarray}
The antisymmetric problem can be treated in a similar manner (see
Appendix \ref{inteqns}).

\subsection{Evaluating ${\tPhi_i}^\pm (\w)$ in the strip   $\Re \w\,\epsilon\,I$}
As already mentioned above, according to Budaev and Bogy's (1995),
evaluation of ${\tPhi}_i^\pm (\w)$ in the strip $\Re
\w\,\epsilon\,I$
 can be carried out by using
the singular convolution type integrals,
 \begin{eqnarray}
\Phi_0^\pm(\w)&=& \frac{1} {4\alpha\ri}\intlimR\frac{f_0^\pm(\eta)}
{\cos\palpha(\frac{\displaystyle \pi}{\displaystyle
  2}-\w+\ri\chi(\eta))}d\eta, \label{tPhi}
\end{eqnarray}
with $f_0^+(\eta)=y^+(\eta)\tanh2\eta
  -\ri c^+_1{\displaystyle \cosh\eta}/{\displaystyle\cosh2\eta}+\ri r_1^+(\eta)\chi'(\eta)$, $f_0^-(\eta)=\ri x^-(\eta)\chi'(\eta)$, $\chi'(\eta)=d \chi/d\eta$
and
\begin{eqnarray} \label{tPsi}
\tPhi_1^\pm(\w)&=&\pm\frac1{4\alpha\ri}\intlimR\frac{f_1^\pm(\eta)\,
  d\eta}{\cos\palpha(\frac{\displaystyle \pi}{\displaystyle
  2}-\w+\ri\eta)},
\end{eqnarray}
with $f_1^+(\eta)=\ri y^+(\eta)$ and
$f_1^-(\eta)=x^-(\eta)\tanh2\eta
  -\ri c^-_1{\displaystyle \cosh\eta}/{\displaystyle\cosh2\eta}-\ri r_1^-(\eta)$.

 If---as is the case for the Sommerfeld amplitudes of the
solution of the original  problem---as $\Imag \w\to \pm\infty$, the
leading terms in (\ref{aspp3}) are
 $O({\rm exp}(\pm \ri p\w))$, with
$\Re p
>0$, then for $\alpha < \pi$,
$-\Re p-\pi/{2\alpha}
 < 0$ and therefore, the
above integrals  converge.

 We start with the
integral of the type (\ref{tPsi}).  Its generic  form is
\begin{eqnarray}
\intlimR \frac{f(\eta)}{\cos \palpha (\xi+\ri \eta)} d\eta, \,\,
|\Re \xi|\le\alpha, \label{S2}
\end{eqnarray}
where the new complex variable is $\xi = \pi/2-\omega$. When $|\Re
\xi| < \alpha$, the integral (\ref{S2})  can be approximated using
the trapezoidal rule. The approximation error is of  order
$O\left[{\rm exp~}(-{2\pi \sigma/ h}) \right]$, with $h$---the
distance between the nodes of a uniform mesh and  $\sigma(\w)$---the
half width of the strip which is centred on the real line and inside
which the integrand is regular. In our case, $\sigma \leq \min
\{\alpha-\rm {Re}~\xi,\alpha+\rm {Re}~\xi \}$ and as $\rm{Re}~\xi
\rightarrow {\alpha}$, the accuracy of the trapezoidal rule
deteriorates.  Therefore, a more robust quadrature formula is
required, with accuracy depending on the function $f(\eta)$ and not
on  parameter $\xi$. One such formula may be obtained with a
modified sinc function,
\begin{eqnarray}
\omega_h(\eta)= \frac{h}{2\alpha}\frac{\sin \frac{\pi}{ h} \eta}{
{\sinh\palpha \eta}}. \label {S3}
\end{eqnarray}
Note that we have
\begin{eqnarray}
\omega_h(nh) = \left\{
\begin{array}{ll}
1,&n=0,     \\
0, \,\,&n \neq 0
\end{array} \right.
 \label{ S4}
\end{eqnarray}
and
\begin{eqnarray}
f(\eta) \approx \sum\limits_{n=-\infty}^\infty
f(nh)\omega_h(\eta-nh) \label {S5}
\end{eqnarray}
where the  sum on the right interpolates $f(\eta)$.  Therefore, the
integral (\ref{S2}) may be approximated by
\begin{eqnarray}
\intlimR {f(\eta)\over \cos\palpha(\xi+\ri\eta)}d\eta\approx
\sum\limits _{-\infty}^\infty {\cal A}_n(\xi)f(nh), \ \label{S6}
\end{eqnarray}
with the coefficients  given by
\begin{eqnarray}
{\cal A}_n(\xi)=\intlimR
 {\omega_h(\eta-nh)\over \cos\palpha(\xi+\ri \eta)}d\eta . \label{S7}
\end{eqnarray}
 These can be evaluated approximately by introducing  new variables
 ${\check \eta}=\eta-nh$ and ${\check \xi}=\xi+\ri nh $.  Then
(\ref{S7}) can be rewritten as
\begin{eqnarray}
&&{\cal A}_n(\xi) = \frac{h}{2\alpha}\intlimR
  {\sin {\pi\over h} {\check \eta}\over \sinh\palpha{\check \eta}}
{d{\check \eta}\over \cos\palpha({\check \xi}+\ri{\check \eta})}=\nonumber\\
&&\frac{h}{4\alpha \ri} \intlimR {\re^ {{\pi\over h}{\check
\eta}\ri} \over \sinh\palpha{\check \eta}} {d{\check \eta}\over
\cos\palpha({\check \xi}+\ri{\check \eta})}- \frac{h}{4\alpha
\ri}\intlimR {\re^{- {\pi\over h}  {\check \eta}\ri}\over
\sinh\palpha{\check \eta}} {d{\check \eta}\over \cos\palpha({\check
\xi}+\ri{\check \eta})}. \label{S7a}
\end{eqnarray}
In the upper (lower) half plane where we can utilize the Jordan
Lemma to  evaluate the first (second) integral, each of the
respective integrands possesses two sets of poles, zeros of $\sinh
\pi{\check \eta}/2\alpha$, $   {\check \eta}=\pm 2 \alpha m\ri,
\,m=0,1,2\dots,$ with the respective residues
\begin{eqnarray}
\pm\frac{2\alpha}{\pi}\frac{\re^{-\frac{2\alpha\pi}{h}m }}{\cos
\palpha{\check \xi}},
\end{eqnarray}
 and  zeros of  $\cos(\pi({\check \xi}+\ri {\check \eta})/(2\alpha))$,
${\check \eta}= \ri [{\check \xi}\pm\alpha(2m+1)],$ with the
respective  residues
\begin{eqnarray}
\mp\frac{2\alpha}{\pi}\frac{\re^{\frac{\pi}{h}[\mp {\check \xi}
-\alpha(2m+1)]}}{\cos \palpha{\check \xi}}.
\end{eqnarray}
It is clear that a significant contribution to (\ref{S7a}) is made
only by the poles, ${\check \eta}=0$ and ${\check \eta}=\ri [{\check
\xi}\pm\alpha]$ (in the upper and lower half plane respectively);
other residues contain small exponential factors. Therefore,
applying the Cauchy Residue Theorem and taking into account that the
first pole lies on the contour of integration, we have
\begin{eqnarray}
\frac{h}{4\alpha \ri}\intlimR {\re^{\pm {\pi\over h}  {\check
\eta}\ri}\over \sinh\palpha{\check \eta}} {d{\check \eta}\over
\cos\palpha({\check \xi}+\ri{\check \eta})}\approx \pm \frac{h}{\cos
\palpha{\check \xi}}\Bigl (  \frac{1}{2} -
\re^{-\frac{\pi\alpha}{h}\mp\frac{\pi}{h}{\check \xi}}\Bigr ),
\label{S28}
\end{eqnarray}
which---returning to the original variables $\xi$ and $\eta$---gives
us a new quadrature formula
\begin{eqnarray}
\intlimR {f(\eta)\over \cos \palpha (\xi+\ri \eta)}d\eta\approx
h\sum\limits_{n=-\infty}^\infty \frac{1}{\cos \palpha(\xi+\ri
nh)}\Bigl [ 1-2(-1)^n&&\re^{-\frac{\pi}{h}\alpha}\cosh {\pi \over
h}\xi
\Bigr ]f(nh),\,\, \nonumber\\
&&|\Re \xi|\le\alpha.\label{new-quad}
\end{eqnarray}

Let us now consider the integrals of the  type (\ref{tPhi}). Its
generic form is
\begin{eqnarray}
\intlimR{\displaystyle {f(\eta)\over\cos\palpha[\xi+\ri \chi
(\eta)]}}d\eta, \,\, |\Re \xi|\le\alpha,  \label{S17}
\end{eqnarray}
where $\chi (\eta)$ is a smooth monotone function. We could change
the integration variable  $\eta$ to $\chi (\eta) $, reduce
(\ref{S17}) to the integral of type (\ref{S2}) and evaluate the
result using a uniform mesh in $\chi$. However, both integrals (\ref
{S2}) and (\ref{S17}) involve  the solution of
(\ref{sing_int_eq_+}), and therefore, it is more reasonable to
evaluate both integrals using the same   mesh. Then following the
same reasoning as above, (\ref{S17}) may be approximated by
\begin{eqnarray}
\intlimR {f(\eta)\over\cos \palpha[\xi+\ri \chi (\eta)]}d\eta
\approx \sum\limits _{-\infty}^\infty {\cal B}_n(\xi)f(nh),
\end{eqnarray}
where the coefficients are given by
\begin{eqnarray}
{\cal B}_n(\xi)=\intlimR
 {\omega_h(\eta-nh)\over \cos \palpha[\xi+\ri \chi(\eta)]}
d\eta , \label{S17a}
\end{eqnarray}
and the main contributions to (\ref{S17a}) are made by the zero
${\check \eta}=0\,(\eta=nh)$ of the hyperbolic sine in $\omega_h$
and the zero ${\check \eta}=\ri a_\pm-nh\,(\eta = \ri a_\pm)$ of the
cosine-function, where $\xi+\ri\chi(\ri a_\pm)=\mp\alpha.$ The
latter equation implies that $ \ri \chi(\ri a_\pm)=-{\rm sin}^{-1}
(\gamma^{-1}\sin a_\pm)=\mp\alpha- \xi, $ and therefore, we have
\begin{eqnarray} a_\pm={\rm sin}^{-1}\bigl[
\gamma\  \sin (\xi\pm\alpha)\bigr], \label {S23}
\end{eqnarray}
with  $\Real a_+>0$ and $\Real a_-<0$. Applying to (\ref{S17a}) the
Cauchy Residue Theorem and noting that the pole ${\check \eta}=0$
lies on the contour of integration we obtain
\begin{eqnarray}
&&\frac{h}{4\alpha \ri} \intlimR {\re^{\pm {\pi\over h} {\check
\eta}\ri}\over \sinh\palpha{\check \eta}} {d{\check \eta}\over
\cos\palpha[\xi+\ri\chi({\check \eta}+nh)] }
\approx\nonumber\\
&&h\Bigl \{ \pm \frac{1}{2\cos\palpha[\xi+\ri\chi(nh)]} \mp \frac{
\re^{\mp\frac{\pi}{h}(a_\pm+\ri nh)}} {{ \chi'}(a_\pm)\sin
\palpha(a_\pm+\ri nh)} \Bigr \},
\end{eqnarray}
where $\chi'(a_\pm)=\cos a_\pm/\sqrt{\gamma^2-\sin^2 a_\pm}.$
Returning to the original variable $\eta$, the resulting  quadrature
formula is
\begin{eqnarray}
&&\intlimR { f(\eta)\over \cos\palpha[\xi+\ri \chi (\eta)] } d\eta
\approx h \sum\limits_{n=-\infty}^\infty \Bigl\{ {1 \over
\cos\palpha[\xi+\ri \chi(nh)]} - \nonumber\\&&(-1)^n \bigl [
\frac{\re^{-\frac{\pi}{h}a_+}}{\chi'(a_+)\sin\palpha(a_++\ri nh)}+
\frac{\re^{ \frac{\pi}{h}a_-}}{\chi'(a_-)\sin\palpha(a_-+\ri nh)}
\bigr ]\Bigl \}f(nh),\nonumber\\&&
\,\,\,\,\,\,\,\,\,\,\,\,\,\,\,\,\,\,\,\,\,\,\,\,\,\,\,\,\,\,\,\,\,\,\,\,\,\,
\,\,
\,\,\,\,\,\,\,\,\,\,\,\,\,\,\,\,\,\,\,\,\,\,\,\,\,\,\,\,\,\,\,\,\,\,\,\,\,\,
\,\,
\,\,\,\,\,\,\,\,\,\,\,\,\,\,\,\,\,\,\,\,\,\,\,\,\,\,\,\,\,\,\,\,\,\,\,\,\,\,
\,\, \,\,\,\,\,\,\,\,\,\,\,\,\,\,\,\,\,\,\,\,\,\,\,\,\,\,|\Re
\xi|\le\alpha. \label{S21b}
\end{eqnarray}
The first terms on the right of both (\ref{new-quad}) and
(\ref{S21b})  effect the trapezoidal rule and the second give a
correction.

\section{Code Testing}
\label{codetesting} Using the above considerations we have developed
a new code for evaluating  the Rayleigh reflection and transmission
coefficients for elastic wedges (see Appendix \ref{coefs}). The
integral equations we solve have the form of the Fredholm equations
of the second kind, but it can be shown that the operators involved
are not Fredholm ({\it cf.}~the statements in Budaev and Bogy 1995,
p.~251). We possess no analytical proof that these equations can be
solved uniquely. Nevertheless, our code produces a solution, and
below we describe verification tests that allow us to state with
confidence that when transformed back to the physical space this
solution satisfies the original diffraction problem. We also
describe successful validation tests, comparing output of our code
with published numerical and experimental data. Of course, the
positive outcomes of these tests do not constitute a theoretical
proof that the code is correct. Note that throughout this section we
characterize materials by their Poisson's ratio $\nu,$ where $
\gamma=\sqrt{(1-2\nu)/[2(1-\nu)]}.$

\subsection{Code verification}

We have designed verification tests to establish  that the computed
functions $\Phi_i(\w)$ are the solutions of the original physical
problem, in particular, that they
\begin{romannum}
\item  are bounded at imaginary infinity;
\item  are analytic  at the boundary of
the strip $\Re \w\,\epsilon\,I$;
\item  possess only  physically
meaningful singularities.
\end{romannum}

The  property (i) is confirmed by direct examination of the computed
 functions
$\widetilde x^-(\eta)$ and $\widetilde y^+(\eta)$ divided by ${\rm
exp}(-|\eta|)$. At large $|\eta|$ the ratios appear to behave as
$O(1)$. It follows that the amplitudes $\Phi_i(\w)$ obtained by the
analytical continuation  must be bounded at the imaginary infinity,
$\eta = \Imag \w \rightarrow \infty.$

\begin{figure}[!ht]
\begin{picture}(200,130)
\put(0,130){(a)} \put(-20,-30){\includegraphics{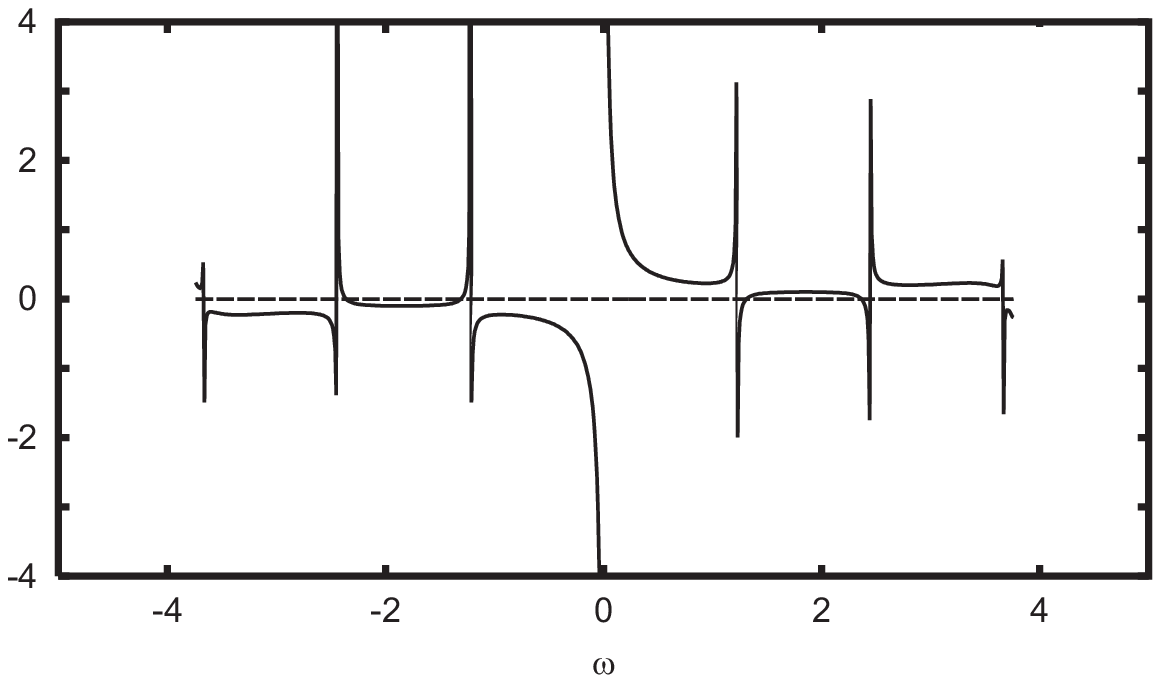}} \put(180,130){(b)}
\put(150,-30){\includegraphics{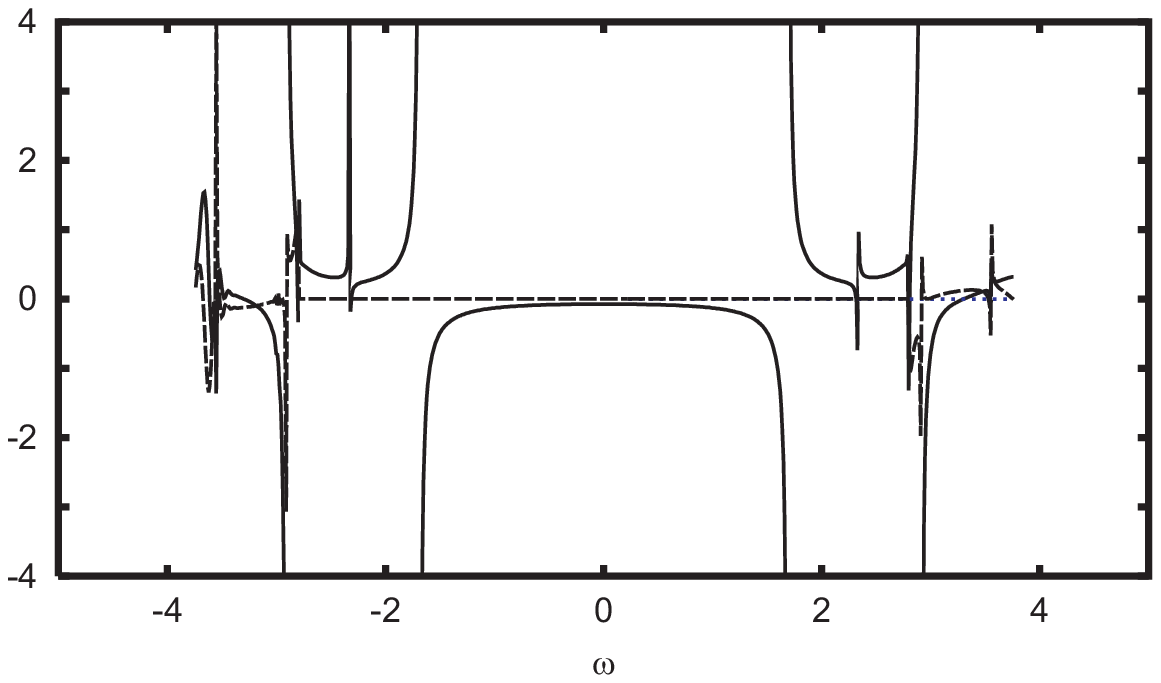}}
\end{picture}
\caption[f]{ The computed Sommerfeld amplitudes: (a) $\Re
\Phi_0(\w)$---dashed line and $\Im \Phi_0(\w)$---solid line, (b)
$\Re \Phi_1(\w)$---dashed line and $\Im \Phi_1(\w)$---solid line.
Wedge angle $2\alpha = 70^{\rm o}$,  $I=[0.96,2.18]$, Poisson's
ratio $\nu=0.25$, incident wave---compressional and
$\theta^{\text{inc}}=0^{\rm o}$.} \label{Sa_70_1s}
\end{figure}

\begin{figure}[!ht]
\begin{picture}(200,130)
\put(0,130){(a)} \put(-20,-30) {\includegraphics{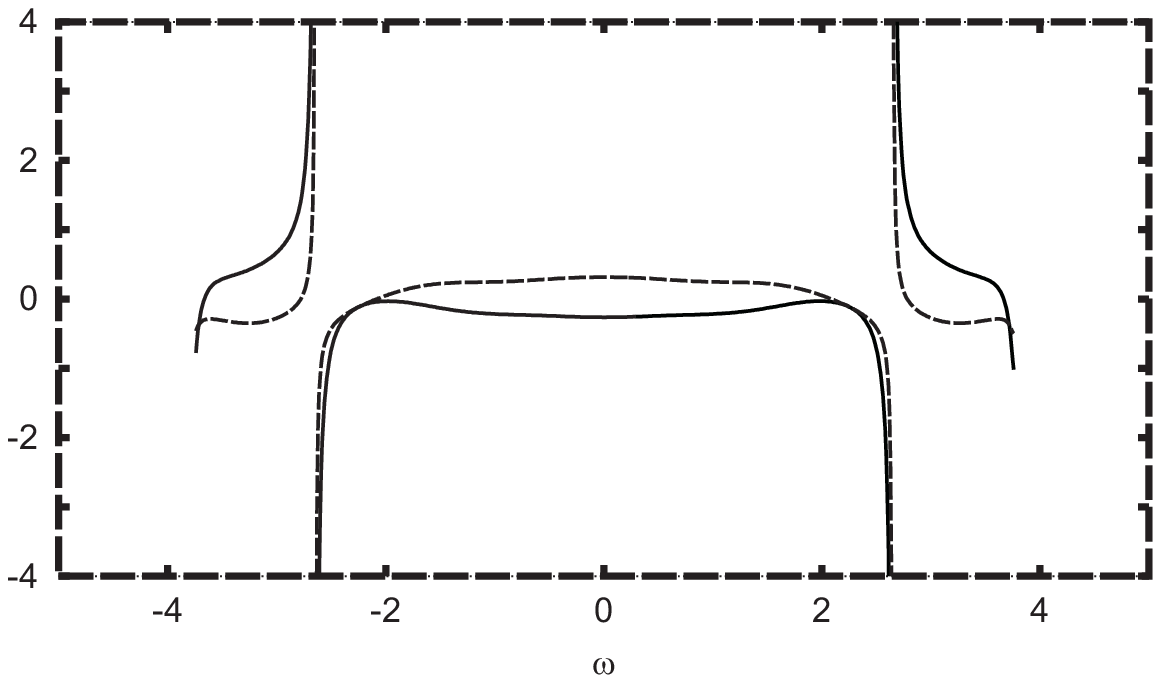}} \put(180,130){(b)} \put(150,-30)
{\includegraphics{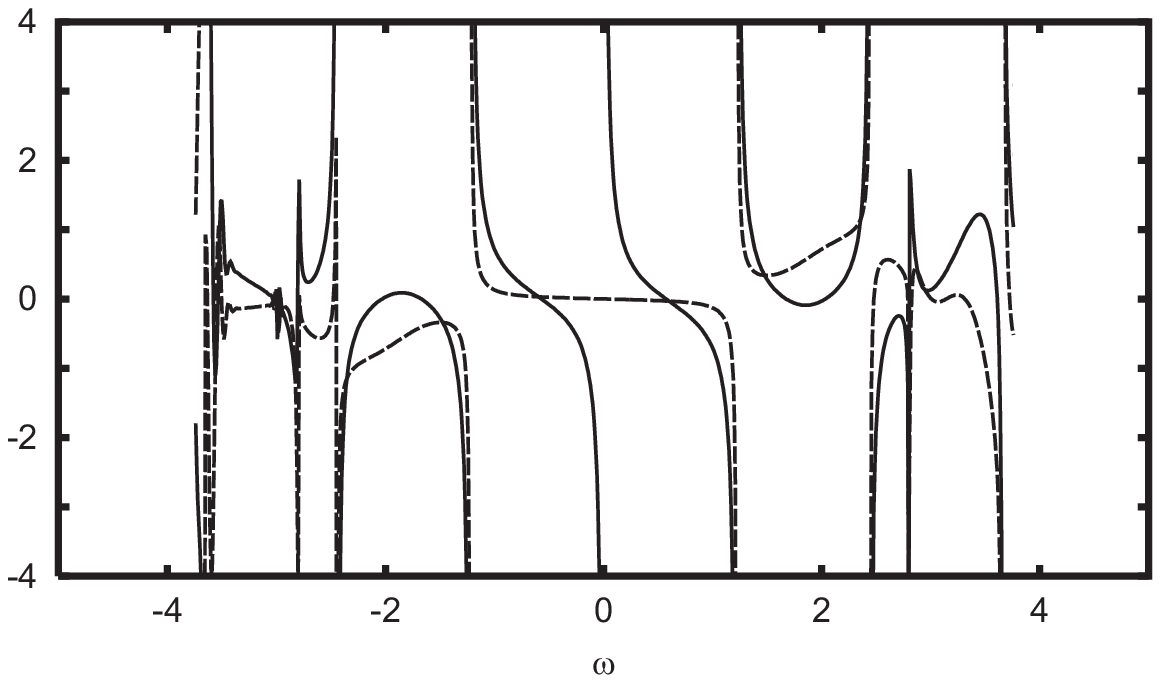}}
\end{picture}
\caption[f]{The computed Sommerfeld amplitudes: (a) $\Re \Phi_0(\w)$
---dashed line and $\Im \Phi_0(\w)$---solid line, (b) $\Re \Phi_1(\w)$
---dashed line and $\Im \Phi_1(\w)$---solid line. Wedge angle
$2\alpha$ = $70^{\rm o}$, $I=[0.96,2.18]$, Poisson's ratio
$\nu=0.25$, incident wave---shear and $\theta^{\text{inc}}=0^{\rm
o}$. }\label{Sa_70_0s} \end{figure}

\begin{figure}[!ht]
\begin{picture}(200,130)
\put(0,130){(a)} \put(-20,-30) {\includegraphics{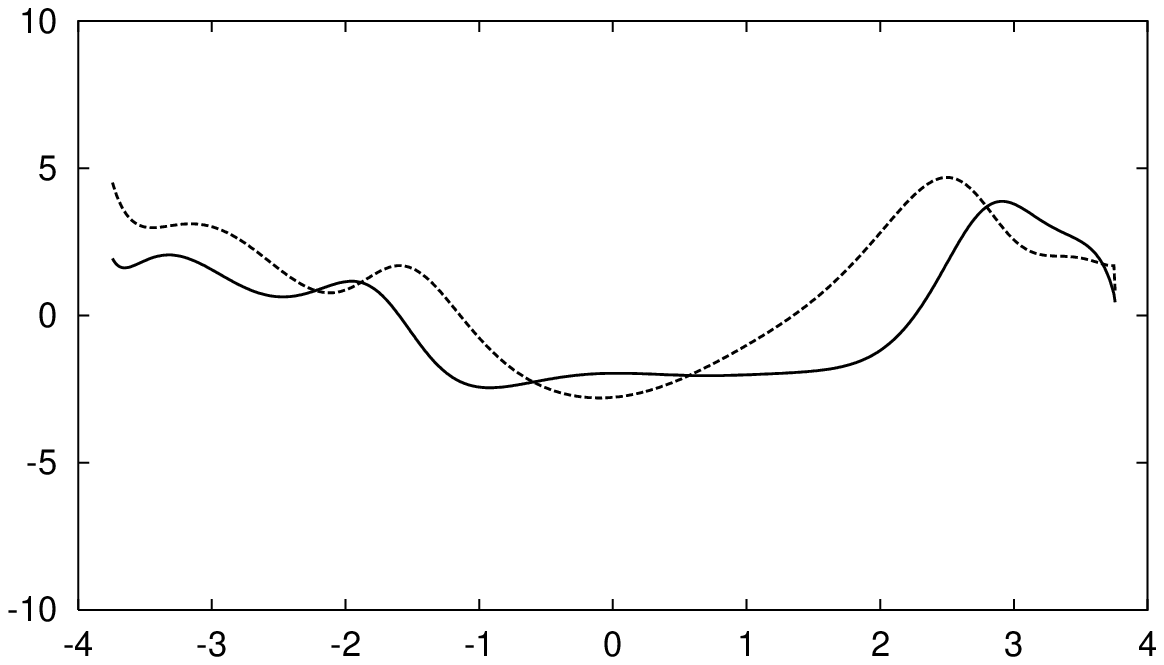}} \put(180,130){(b)} \put(145,-30)
{\includegraphics{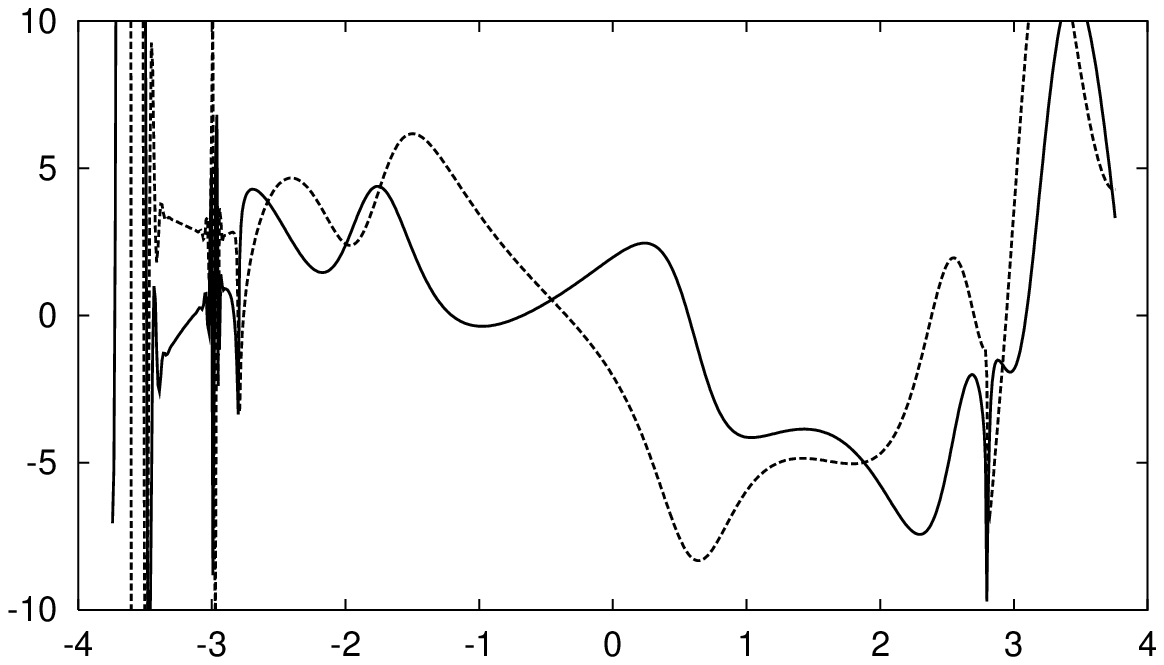}}
\end{picture}
\caption[f]{The computed Sommerfeld amplitudes: (a) $\Re \Phi_0(\w)$
---dashed line and $\Im \Phi_0(\w)$---solid line, (b) $\Re \Phi_1(\w)$
---dashed line and $\Im \Phi_1(\w)$---solid line.
  Wedge angle
$2\alpha$ = $70^{\rm o}$,  $I=[0.96,2.18]$, Poisson's ratio
$\nu=0.25$, incident wave---Rayleigh and $\Re
\theta^{\text{inc}}=-\alpha$.} \label{Sa_70_ray}
\end{figure}

Since the last step in the analytical continuation is carried out
strip by strip, all  $2\alpha$ wide, there is no guaranty that any
computed $\Phi_i(\w)$ should be smooth at the boundaries of the
initial strip $\Re \w\,\epsilon\,I$. However, examination of the
numerical output related to Figs. \ref{Sa_70_1s}-- \ref{Sa_70_ray}
confirms that our approximations {\it are}
 smooth:
The  numerical derivatives of computed $\Phi_i(\w)$ jump at the
boundaries of the strip by about $10^{-4}$. It follows that the
property (ii) is satisfied. Interestingly, when attempting to solve
an incorrect problem, with the constants $c_1^\pm$ put to zero,  the
computed $\Phi_i(\w)$ themselves  jump at the boundaries by about
$10^{-2}$.

Similarly to (ii), the property (iii) should be satisfied by the
Sommerfeld amplitudes of the solutions of the original wedge
diffraction problem, but it is not obvious that the computed
solutions of the corresponding functional equations should satisfy
it as well. Indeed, the way they are constructed assures that the
computed amplitudes $\Phi_i(\w)$ have physically meaningful poles
and since the functional equations that are used to effect the
analytical continuation involve reflection coefficients ${\rm
r}_{jk}(\w)$, with the branch point at $\w=\theta_h$ and poles at
$\w=\pm \ri\beta_{\rm R}$ (see (\ref{san_cont}) and Appendix
\ref{nomenclature}), they possess physically meaningful branch
points and Rayleigh poles too. However, by the same token, the
analytical continuation scheme {\it might}  endow these amplitudes
with {\it extra, physically meaningless singularities}. Remarkably,
when the incident wave is compressional or shear at $\w=-\alpha\pm
\ri\beta_{\rm R}$  all our computed residues are  of order
$10^{-7}$, i.e.~are numerical zeros. Thus, the computed Sommerfeld
amplitudes possess no Rayleigh poles corresponding to physically
meaningless Rayleigh waves incoming from infinity. Furthermore, Fig.
\ref{Sa_70_1s}--\ref{Sa_70_0s},  which respectively relate to a
purely symmetric and purely asymmetric case, confirm that inside a
neighborhood of zero which includes  the strip $\Re
\w\,\epsilon\,I$, the computed Sommerfeld amplitudes possess the
symmetries described in (\ref{sas_deco}), that is $\Phi_0^+(\w)$ and
$\Phi_1^-(\w)$ are odd while $\Phi_1^+(\w)$ and $\Phi_0^-(\w)$  are
even. (Outside this neighborhood, the symmetries are not apparent in
Figs.~\ref{Sa_70_1s}-- \ref{Sa_70_ray} due to the accumulation of
numerical errors.) As we  show in Appendix \ref{app:Rploes_brcuts},
such symmetries imply the absence of physically meaningless branch
points.

The properties (i), (ii) and (iii) of the computed Sommerfeld
amplitudes respectively imply that they possess all the properties
expected of the Sommerfeld amplitudes of the solutions $\phi_i(kr,
\theta)$ of the original diffraction problem, so that their
corresponding Sommerfeld integrals satisfy (i) the Helmholtz
equations and correct tip condition; (ii) zero stress boundary
conditions; and (iii)  radiation conditions  (which exclude
nonphysical Rayleigh or head  waves incoming from infinity).

Fig. \ref{Sa_70_1s} provides one more confirmation that the computed
functions $\Phi_i(\w)$ are the Sommerfeld amplitudes of solutions
$\phi_i(kr, \theta)$  of the original wedge problem: They show that
for the  symmetric  compressional wave incidence, both $\Phi_i(\w)$
are imaginary and therefore, the corresponding displacements are
real. This is consistent with the physics of the problem, since
unlike with the symmetric  shear wave incidence, there is no total
internal reflection, that is no imaginary displacement component.

Finally, a  numerical stability of the scheme is ascertained   by
the fact that different choices of adjustable function $\sigma(\w)$
in (\ref{stard}) all give similar results (see Appendix
\ref{numerics}).

\subsection{Code validation}

Our first validation results are presented in  Tables
\ref{RS_fujii_us1} and \ref{RS_fujii_us2}, where  the approximate
values of amplitudes and phases of reflection  and transmission
coefficients  $R^{\rm ref}$ and $R^{\rm tran}$  as computed with our
code are compared with numerical results of Fujii (1994). Each
Fujii's column contains  values corresponding to different choices
of an adjustable parameter.  The parameter allows one to evaluate
singular integrals on the real axis by moving the poles away from
the axis into the complex plane.  This is equivalent to employing
the radiation condition at infinity in the form of the limiting
absorption principle. From the physical point of view, the
singularities cannot be moved too far.  However, when they are too
close the evaluation algorithm becomes numerically unstable. The top
rows in the tables are obtained with the parameter values that
correspond to a more physically meaningful situation and the bottom
ones, with the  values that give  better numerical stability. The
tables demonstrate that for the larger wedge angles the agreement
with our computations is quite good, but for the smaller ones our
values lie outside Fujii's range. This is not surprising, because
when the wedge angles are small there are many multiply reflected
waves and the residues of  many resulting  poles   are large. For
this reason, when the wedge angles are small, the present version of
our code looses its numerical stability.

\begin{figure}[!ht]
\begin{picture}(200,130)
\put(0,130){(a)} \put(-20,-30) {\includegraphics{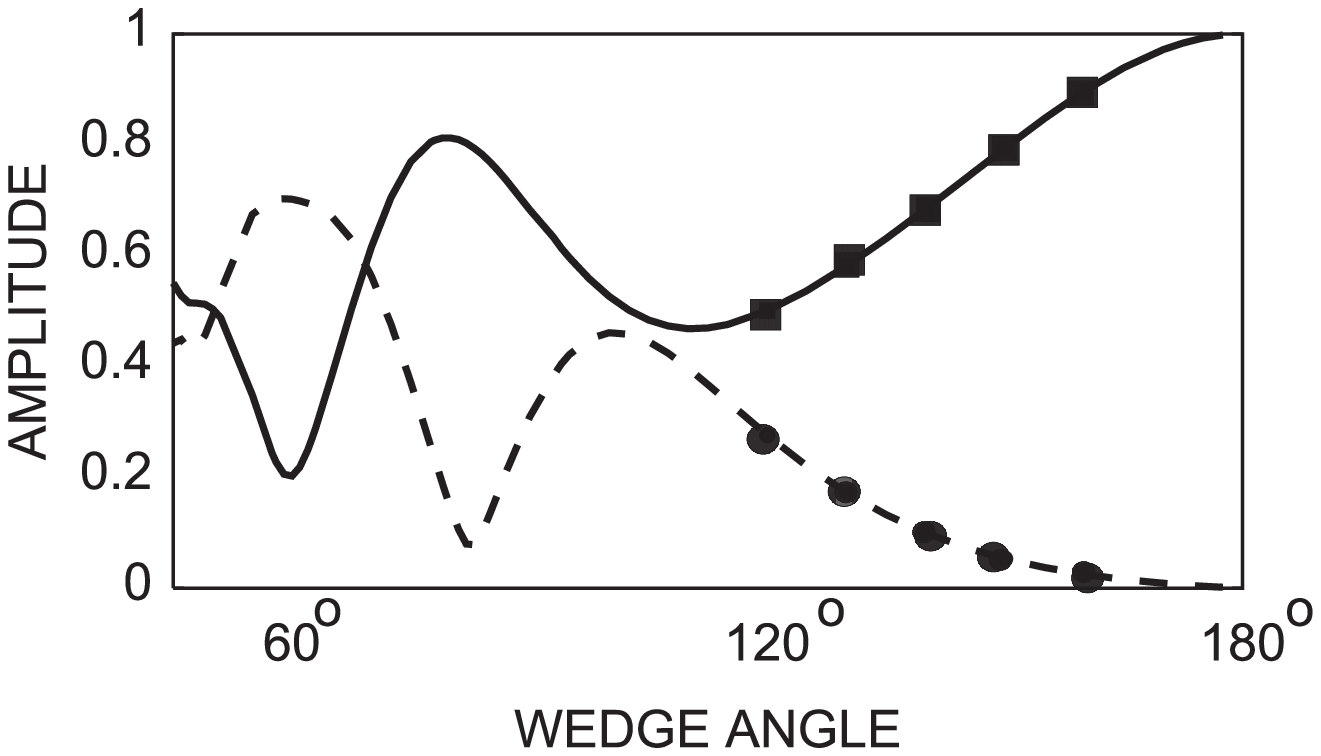}} \put(180,130){(b)} \put(170,-30)
{\includegraphics{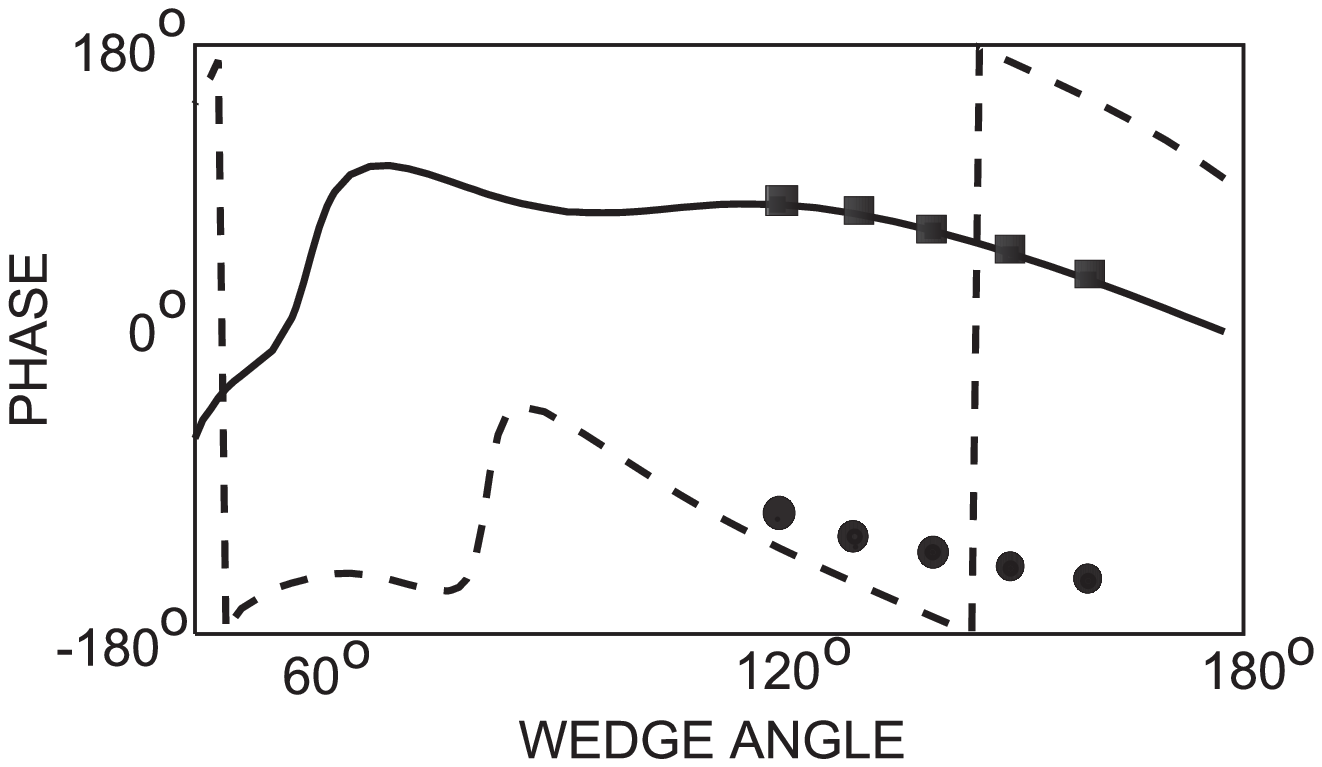}}
\end{picture}
\caption[f]{Rayleigh transmission coefficients (solid line) and
reflection coefficients (dashed line)  computed with our code versus
the coefficients computed with Budaev and Bogy's code (squares and
circles respectively---see Budaev and Bogy 2001, Fig. 6). Poisson's
ratio $\nu =0.234$, incident wave---Rayleigh and $\Re
\theta^{\text{inc}} = -\alpha$.} \label{rs_0234_av}
\end{figure}

\begin{figure}[h]
\begin{picture}(200,130)
\put(0,130){(a)} \put(-130,-480) {\includegraphics{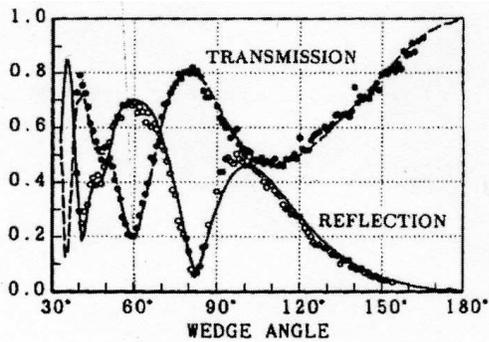}} \put(200,130){(b)} \put(52,-485)
{\includegraphics{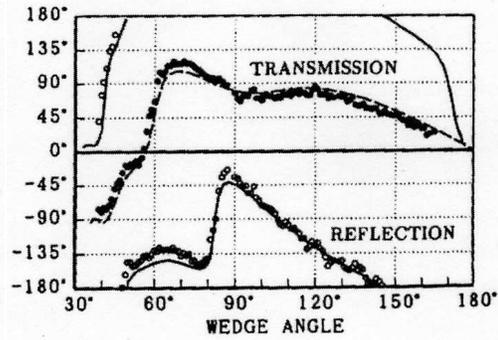}}
\end{picture}
\caption[f]{The Fujii's computed (solid line) and experimental
(dots) transmission and reflection  coefficients: (a) amplitudes,
(b) phases.  Poisson's ratio $\nu =0.234$, incident wave - Rayleigh
and $\Re \theta^{\text{inc}} = -\alpha$. Reproduced from Fig. 7 in
Fujii (1994).}\label{fujii_gr}
\end{figure}

To continue, in Figs.  \ref{rs_0234_av}
 (a) and (b)  we present
our Rayleigh  reflection and transmission coefficients  as functions
of the wedge angle, computed  for $\nu = 0.234$. They fit Fujii's
numerical and experimental data extremely well (see our Fig.
\ref{fujii_gr} or Fujii 1994, Fig. 7). Note that on Fujii's plots
the solid lines represent his numerical results and  discrete
points, his experimental data.) Indeed, for the wedge angles between
$45^o$ and $150^o$  we cannot put the  results on the same
graph---there is no visible difference. Note  that the jumps in the
phase of the reflection coefficient that take place at the wedge
angles of about $45^{\rm o}$  and $145^{\rm o}$ are from $180^{\rm
o}$ to $-180^{\rm o}$ and $-180^{\rm o}$ to $180^{\rm o}$
respectively, and therefore, no jumps in physical quantities take
place. For the wedge angles between $150^{\rm o}$ and $180^{\rm o}$
the reflection coefficients are practically zero. This is
understandable, because when the wedge angle is $180^{\rm o}$ there
is no reflection. In this region, the phases of our reflection
coefficients differ from Fujii's but the limiting value of
$90{}^{\rm o}$ agrees with the one obtained by Gautesen (2002a). It
appears that in this region Fujii's scheme looses its stability.

\begin{table}[htbp]
\caption{Rayleigh reflection coefficients computed  with our code
and Fujii' (see Fujii 1994). $\nu = 0.25$.}\label{RS_fujii_us1}
\begin{center}\footnotesize
\renewcommand{\arraystretch}{1.3}
    \begin{tabular}{c c c c c c c c c}
  \hline
wedge        &\multicolumn{2}{c}{$|R^{\rm ref}|$} &
\multicolumn{2}{c}{$\text{arg }R^{\rm ref}$}
\\
  angle      & Fujii  & This paper & Fujii  & This paper
\\ \hline
             & 0.50552 &             &  $-169.87^{\rm o}$        \\
 $50^{\rm o}$& 0.49924 &  &  $-169.54^{\rm o}$ &        \\
             & 0.49278 &0.47427  &  $-169.07^{\rm o}$ & $-161.4^{\rm o}$             \\ \hline
             & 0.05257 &             &  $170.53^{\rm o}$      \\
             & 0.05252 &             &  $170.14^{\rm o}$        \\
$150^{\rm o}$& 0.05236 &0.05197      &  $169.85^{\rm o}$ & $170.5^{\rm o}$  \\
             & 0.05217 &             &  $169.65^{\rm o}$ \\
             & 0.05196 &             &  $169.53^{\rm o}$  \\
             & 0.05151 &             &  $169.43^{\rm o}$  \\
   \hline
\end{tabular}
\end{center}
\end{table}

\begin{table}[htbp]
  \caption{Rayleigh transmission coefficients computed  with our code and
Fujii' (see Fujii 1994). $\nu = 0.25$.}\label{RS_fujii_us2}
\begin{center}\footnotesize
\renewcommand{\arraystretch}{1.3}
    \begin{tabular}{c c c c c c c c c}
  \hline
wedge & \multicolumn{2}{c}{$|R^{\rm tran}|$} &
\multicolumn{2}{c}{$\text{arg }R^{\rm tran}$}
\\
  angle      & Fujii  & This paper
 & Fujii  & This paper
\\ \hline
             & 0.49123     & & $-32.59^{\rm o}$ &                 \\
$50^{\rm o}$ & 0.48372     & & $-32.84^{\rm o}$ &                 \\
             & 0.47552     & 0.55189 & $-33.00^{\rm o}$ &$-26.9^{\rm o}$            \\ \hline
             & 0.78940     & & $52.79^{\rm o}$   &         \\
             & 0.78899     & & $52.80^{\rm o}$   &         \\
$150^{\rm o}$ & 0.78866 & 0.78942 & $52.81^{\rm o}$   &  $52.9^{\rm o}$   \\
             & 0.78842     & & $52.82^{\rm o}$  &         \\
             & 0.78825     & & $52.84^{\rm o}$   &         \\
             & 0.78800     & & $52.86^{\rm o}$   &         \\
   \hline
\end{tabular}
\end{center}
\end{table}

\begin{figure}[!ht]
\begin{picture}(200,100)
\put(10,100){(a)} \put(-20,-20)
 {\includegraphics{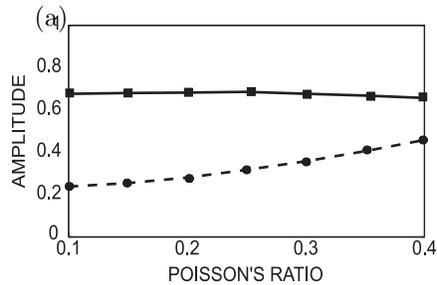}} \put(190,100){(b)} \put(180,-20)
{\includegraphics{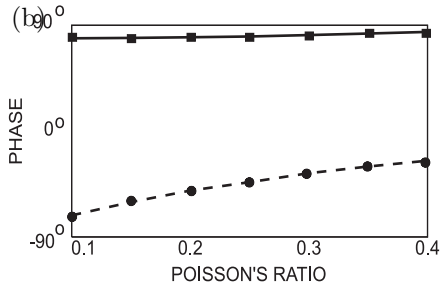}}
\end{picture}
\caption[f]{Rayleigh transmission  coefficients (solid line and
squares) and reflection coefficients (dashed line and circles)
computed respectively with our code and Gautesen's code (see
Gautesen 2002a, Figs. 3 and 4): (a) amplitudes, (b) phases. Wedge
angle $90^{\rm o}$, incident wave---Rayleigh and $\Real
\theta^{\text{inc}} = -\alpha$.} \label{ray90}
\end{figure}

The amplitude curves reported by Budaev and Bogy (2001) are the same
as ours (see {\it ibid}, Fig. 6 and our Fig. \ref{rs_0234_av} (a)),
but for the larger wedge angles, the phase of their reflection
coefficient is somewhat different---see Fig. \ref{rs_0234_av} (b).
The discrepancy might not be  crucial, because at these angles  the
amplitudes of the reflection coefficients are very small, but the
problem is indicative of numerical instability. Note that the
results on the wedge
 angles
 greater than $180^{\rm o}$ as presented by Budaev and Bogy (1996) are
incorrect---see their errata (Budaev and Bogy 2002).   Note too that
even though Poisson's ratio used by Budaev and Bogy (2001) is $\nu =
0.294$ the above comparison is valid: The coefficients should not be
effected by a small difference in $\nu$ (see e.g.~Fig. \ref{ray90}.)

We finish this section by comparing our computed Rayleigh reflection
and transmission coefficients for the quarter space with Gautesen's.
On taking into account that Gautesen's coefficients are complex
conjugates of ours and thus, our phases must have the opposite sign,
the agreement between the calculations is very good (see
Fig.~\ref{ray90}).


\section{Conclusions}

We have studied the properties of the underlying integral operators
and developed a new numerical schedule for solving the singular
integral problem that arises in   Budaev and Bogy's approach to
diffraction by two dimensional traction free isotropic elastic
wedges. We have also developed new quadrature formulas for
evaluating the singular convolution type integrals that are utilized
in this approach. Although the analytical justification of the
method is not entirely rigorous, the code has undergone a series of
stringent internal verification tests directed at establishing that
it solves the original physical problem as well as validation tests
against numerical and experimental results reported by other
authors. It appears to be successful when simulating diffraction by
wedges of angles between $40^{\rm o}$ and $178^{\rm o}$ of  plane
incident waves, compressional or shear. When the incident wave is a
Rayleigh the lower limit of applicability goes up to $45^{\rm o}$.

\clearpage
\appendix
\section{Analytic properties and asymptotic expansions of $\Phi_i$} \label{app:el_pot_repr}

In this Appendix we first follow Malyuzhinets (1957), Osipov and
Norris (1999) and Budaev and Bogy (1995) and explain in more detail
why the elastic potentials $\phi_i$  may be represented in the form
of the Sommerfeld integrals. We then elucidate the analytic
properties and asymptotic expansions of functions $\Phi_i(\w)$.

\subsection{The representation of $\phi_i$  in the form of the
Sommerfeld integral}

 We develop our arguments  for the potential $\phi_1$ only.
The potential $\phi_0$ may be treated in a similar manner.   Let us
introduce the Laplace transform
\begin{eqnarray}
\overline\Phi_1(s,\theta)=k\intlim_0^{\infty} \psi(kr,\theta)
\re^{-ksr}\rd r. \label{lapl}
\end{eqnarray}
 The behaviour of $\psi(kr,\theta)$ at infinity is known from the
physics of the problem: The potential comprises the incident and
reflected waves, waves diffracted by the wedge tip, head and
Rayleigh waves. As $r\rightarrow \infty$ the corresponding terms in
the asymptotic expansion of the function $\psi $ are of order
$O(1)$. Their derivatives  in both $r$ and $\theta$ are of the same
order.

  It follows  that the function $\overline\Phi_1(s,\theta)$
is defined and regular on the whole half-plane  $\rm{Re}~s>0$.  As
$s\rightarrow \ri\tau+0$, where $\tau$ is real, the function
$\overline\Phi_1$ becomes the Fourier transform of $\psi(kr,\theta)$.
For any $\theta$, the function $\overline\Phi_1(\ri\tau+0,\theta)$ has
a finite number of singularities. Each  is associated with one of
the waves mentioned above.  Excluding these singularities, the
function $\overline\Phi_1(\ri\tau+0,\theta)$ is an analytic
 function of $ \tau$.

\begin{figure}[!ht]
\centering
\includegraphics[scale=0.5]{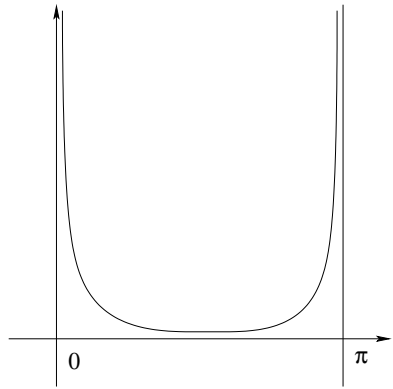}
\caption{ The Sommerfeld contour.} \label{som_contour}
\end{figure}

Using the inverse Laplace transform, $\psi(kr,\theta)$ may be
written as
\begin{equation}
\psi(kr,\theta)=\frac1{2\pi\ri}\intlim_{\sigma-\ri\infty}^{\sigma+\ri\infty}
\overline\Phi_1(s,\theta)\re^{ kr s}\rd s. \label{laplace}
\end{equation}
Introducing a new independent variable $\w$, such that
$s=\ri~\cos\w$, Eq. (\ref{laplace}) becomes the Sommerfeld integral
\begin{equation}
\psi(kr,\theta)=\intlim_C \hat\Phi_1(\w,\theta)\re^{\ri k r \cos\w}\rd
\w, \label{somm_integral}
\end{equation}
where we use the notation
\begin{equation}
\hat\Phi_1(\w,\theta)={1\over 2\pi}\sin\w~
\overline\Phi_1(\ri~\cos\w,\theta), \label{change_var}
\end{equation}
and $C$ is a $\bigcup$ shaped Sommerfeld contour running from
$\ri\infty$ to $\pi+\ri\infty$ (see Fig.~ \ref{som_contour}).

Since $\overline\Phi_1(s,\theta)$  is regular in the right
 half-plane ${\rm Re~}s>0$
 and the new variable $s=i\,\cos \w$  maps this
half-plane onto the half-strip $\{0 < {\rm Re~} \w < \pi,\ \Imag
\w>0\}$, it follows that $\hat\Phi_1(\w, \theta)$  is regular inside
this half-strip.

\subsection{The analytic
properties of $\hat\Phi_1$}

The tip conditions assure that as $kr\rightarrow  0$,
$\psi(kr,\theta)$ and derivatives $\partial_\theta\psi$, and
$\partial^2_\theta \psi $ remain bounded. The boundedness of $\psi$
implies that
 for a fixed $\arg s $, such that
$-{\pi/2}<\arg s<{\pi/2}, $ the function $\overline\Phi_1(s,\theta)=
O(s^{-1})$.  Taking into account the formula (\ref{change_var}) this
in its turn implies that $\hat\Phi_1(\w,\theta)=O(1) $.  The same
estimate is valid for the derivatives  of $\hat\Phi_1$, so that
$\partial_\theta \hat\Phi_1, \
\partial ^2_ \theta\hat\Phi_1 = O(1)$.

Let us assume  that
\begin{remunerate}
\item excluding a finite number of poles and branch points
 on the boundary of the  half-strip $\{0<\rm{Re}~ \w<\pi,\ \rm{Im}~\w>0\}$, for
any fixed $\theta$, the function  $\hat\Phi_1$ is regular inside the
half-strip
\begin{eqnarray}
-\varepsilon < \rm {\Re}~\w < \varepsilon+ \pi, \    \rm{Im}~\w
>-\varepsilon ; \label {wide}
\end{eqnarray}

\item $\hat\Phi_1$ is regular in the vicinity of $0$;

\item   for large $\rm
{Im}~\w$, inside the half-strip (\ref {wide}), $ \hat \Phi_1,\
{\partial_\theta }\hat\Phi_1, \ {\partial ^2_\theta\theta}\hat \Phi_1 =
O(1).$
\end{remunerate}
These assumptions  may be justified  using results of Kamotski and
Lebeau (2006);  they allow us to transform the contour $C$ (see
Fig.~ \ref {som_contour}) into the contour
\begin{eqnarray}
C_{\varepsilon^\prime }= \left\{
\begin{array}{rll}  \rm {Re}~\w&=-\varepsilon^\prime,&\  \Imag \w\geq \rm
{const},\\ -\varepsilon^\prime &\leq \Re \w\leq
\pi+\varepsilon^\prime,&\  \Imag \w =\rm {const},\\ \rm {Re}~\w&=
\pi+\varepsilon^\prime,&\ \Imag \w\ge \rm{const,}
\end{array}\right.   \label{wide_contour}
\end{eqnarray}
 where the  constant  is  sufficiently large  and
$0<\varepsilon^\prime<\varepsilon$.

Following Budaev (1995) let us transform the contour $C$   in the
integral (\ref{somm_integral}) into the contour
$C_{\varepsilon^\prime}$ and substitute the resulting integral into
the Helmholtz equation. We
 obtain
\begin{equation}
\intlim_{C_{\varepsilon^\prime}
}\Big(\partial^2_\w-\partial^2_\theta\Big)\hat\Phi_1(\w,\theta)\re^{k\ri
r \cos\w} \rd \w =0. \label{A5}
\end{equation}
The Nullification Theorem implies that inside the contour
$C_{\varepsilon^\prime}$ we have
\begin{eqnarray}
\Big(\partial^2_\w-\partial^2_\theta\Big)\hat\Phi_1(\w,\theta)=0.
\label{string}
\end{eqnarray}
Excluding the poles and branch points, the function $\hat\Phi_1$ and
its derivatives are regular functions inside the contour
(\ref{wide_contour}).
 Therefore,
(\ref{string}) is valid inside the region (\ref {wide}).

Eq. (\ref{string}) implies
\begin{equation}
\hat\Phi_1(\w,\theta)=\Phi_1^{(1)}(\w+\theta)+\Phi_1^{(2)}(\w-\theta).
\label{dalam}
\end{equation}
Let us apply the operator $\partial_\w +\partial_\theta $ to both
sides of (\ref{dalam}). This gives
\begin{eqnarray}
 \left({\partial_\w}
+{\partial_\theta }\right)\hat\Phi_1 (\w,\theta)=
 2[\Phi_1^{(1)}]^\prime (\w+\theta).
\label{anal}
\end{eqnarray}
Excluding its poles and branch points, the function in the left-hand
side of (\ref{anal}) is  regular  inside the region
 $\{|\theta|\leq \alpha,\  \Imag \w> -\varepsilon,\
-\varepsilon< \Re~ \w< \pi+\varepsilon\}$. Therefore, both $\Phi_1^{(1)}$
and $[\Phi_1^{(1)}]^\prime$   are regular inside this region and excluding
its poles and branch points, $\Phi_1^{(1)} (\w)$  is regular in the
half-strip
\begin{eqnarray}
-\alpha -\varepsilon < \rm \Real \w < \varepsilon+\alpha+ \pi, \
\rm{Im}~\w
>-\varepsilon. \label{anal1}
\end{eqnarray}
Applying the operator ${\partial_\w} -{\partial_\theta } $ to both
sides of (\ref{dalam}) we obtain
\begin{eqnarray}
 \left({\partial_\w}
-{\partial_\theta }\right)\hat \Phi_1 (\w,\theta)=
 2[\Phi_1^{(2)}]^\prime (\w-\theta).
\label{anal2}
\end{eqnarray}
Eq.  (\ref{anal2})  implies regularity of $\Phi_1^{(2)}(\w)$  in the
half-strip
\begin{eqnarray}
-\alpha -\varepsilon-\pi < \Re\w < \varepsilon+\alpha, \    \Imag \w
<\varepsilon. \label{anal3}
\end{eqnarray}
As $kr\rightarrow \infty$,  regularity of $\Phi_1^{(1)}$ and $\Phi_1^{(2)}$
allows us to
 evaluate
 the integral
(\ref{somm_integral}) using the stationary phase method.   The
stationary points are $\w=0$ and $\w=\pi$.
 The asymptotic term
$ \Phi_1(0,\theta){\rm exp}~[\ri (kr+\pi/4)]/\sqrt{2\pi k r},$
evaluated at $\w=0$ is zero, because there can be no cylindrical
wave incoming from infinity.  Therefore, we have
\begin{equation}
\hat\Phi_1(0,\theta)=\Phi_1^{(1)}(\theta)+\Phi_1^{(2)}(-\theta)=0,
\end{equation}
and in the neighbourhood of $ \theta=0$,
\begin{eqnarray}
\Phi_1^{(2)}(\theta)=-\Phi_1^{(1)}(-\theta) \label{anal4}.
\end{eqnarray}
It follows that (\ref{anal4}) is satisfied identically.  Therefore,
we have
\begin{eqnarray}
\hat\Phi_1 (\w,\theta)=\Phi_1^{(1)}(\w+\theta)- \Phi_1^{(1)}(-\w+\theta) \label
{anal5}
\end{eqnarray}
and
\begin{eqnarray}
 \psi(kr, \theta) = \intlim_{C_{\varepsilon^\prime}}
[\Phi_1(\w +\theta)-\Phi_1(-\w+\theta)] \re^{k\ri r\cos\w} ~ \rd\w,
\label {anal6}
\end{eqnarray}
where we dropped the superscript $(1)$. A further change in notations
finally gives us Eq. (\ref{S2int3}).

\begin{figure}[!ht]
\centering
\includegraphics[scale=0.5]{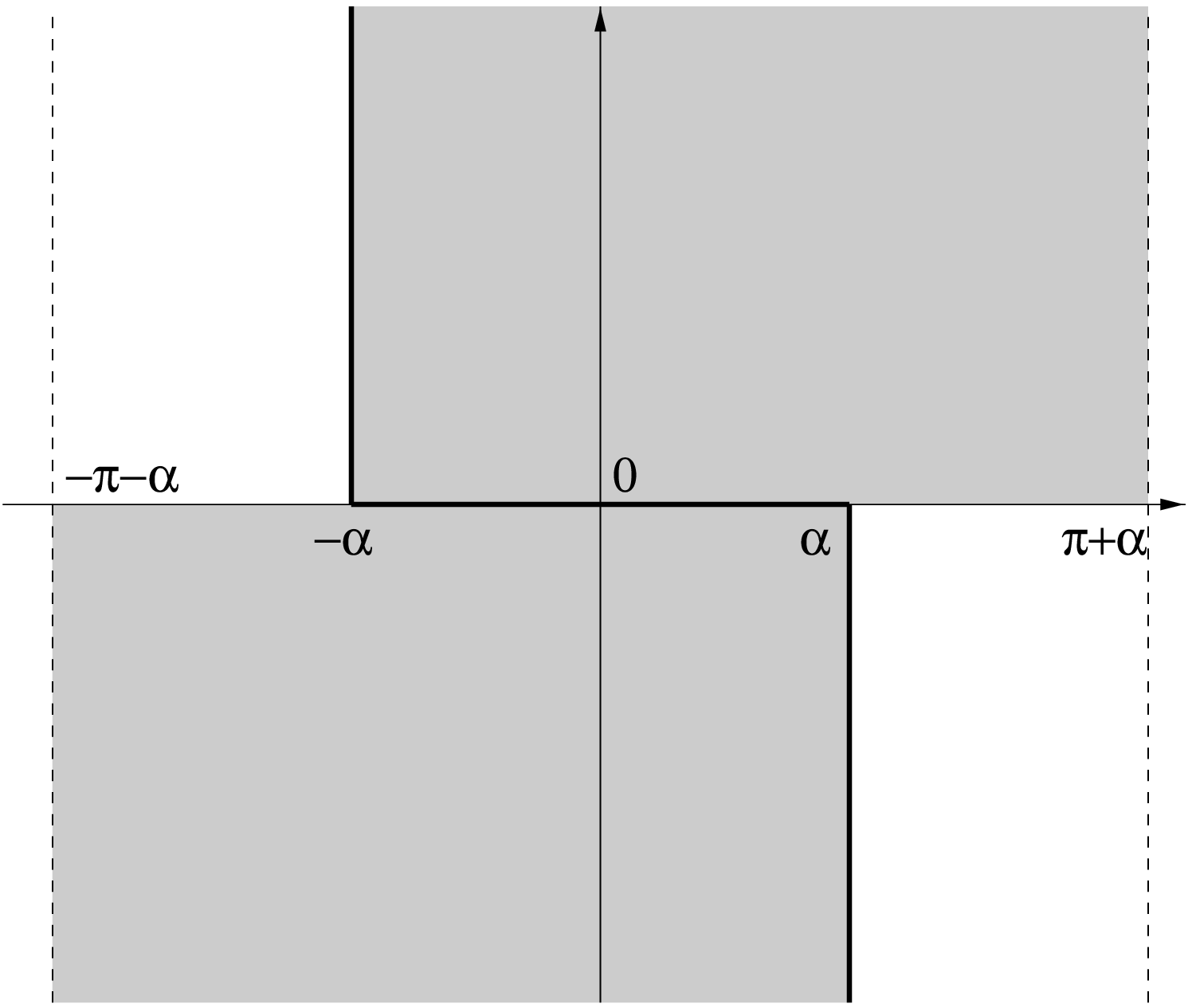}
\caption{ The regularity domain of $\Phi_1(\w)$.} \label{reg_domain}
\end{figure}

\subsection{The domain of regularity and location of
singularities  of $\Phi_1 (\w)$ }

Let us return to (\ref {anal}).
 Its left-hand side  is regular in the region $\{\Imag \w>0,\ 0<\Re\w<\pi\}; |\theta|\leq \alpha\}$. Therefore, the function  $\Phi_1(\w)$ has no
singularities in the half-strip $\{-\alpha<\Re \w<\pi+\alpha,\ \Imag
\w>0\}$. It can be similarly argued that the half-strip
$\{-\pi-\alpha<\Re \w<\alpha,\ \Imag \w<0\}$ contains no
singularities either
 (see Fig.~ \ref{reg_domain}, where the dashed
and solid lines are the boundaries of the two half-strips in which
the function $\Phi_1(\w)$ is holomorphic.) However, there might be
singularities on the boundaries of these half-strips.

Let us  consider the solid  lines
\begin{eqnarray}
\{\Re\w=-\alpha,\ \Imag \w\geq0\}\cup\{\Imag \w=0,\
-\alpha\leq\omega\leq +\alpha,\}\ \cup\{\Re\w=\alpha,\ \Imag
\w\leq0\}.
\end{eqnarray}
The
 singularities which lie on these lines give rise to  incoming surface
 waves. Let us discuss this point in more detail. Let us consider
 the Sommerfeld integral
\begin{eqnarray}
\intlim_{C\cup \tilde C}\Phi_1(\w+\theta) \re^{\ri kr\cos \w}\rd\w.
\end{eqnarray}
When evaluating (\ref{S2int3}), any pole $\w_0$ of $\Phi_1(\w)$, such
that $ - \alpha<\w_0<\alpha,$ gives rise to a plane wave with the
phase factor
 ${\exp}~[\ri kr\cos (\w_0-\theta)]$ (see (\ref{anal7})), that is  a plane {\it body} wave
 propagating  along the ray
 $\theta = \w_0$ which lies inside the  wedge. By the same token, any pole
$\w_1=-\alpha+\ri \beta_R$ of $\Phi_1(\w)$, with $\beta_R>0$ gives
rise to a plane wave with the phase factor
\begin{eqnarray}
\re^{\ri kr \cos (\w_1-\theta)}= \re^{\ri kr \cos (\alpha+\theta
)\cosh \beta_R}\re^{-kr{\sin(\alpha+\theta) \sinh \beta_R}},
\label{anal7}
\end{eqnarray}
so that its amplitude is exponentially small everywhere except for a
small neighbourhood of the wedge face $\theta = -\alpha$. Thus, the
pole $\w_1=-\alpha +\ri\beta_R,\ \beta_R>0$ gives rise to the {\it
surface} (known as Rayleigh) wave which  propagates from infinity
to the wedge  tip along the wedge face $\theta=-\alpha$.  Similarly,
any pole $\w_2=\alpha-\ri\beta_R$ corresponds to the Rayleigh {\it
surface} wave which
 propagates
from infinity  to the wedge tip along the wedge face
$\theta=\alpha$.

\subsection{The  asymptotics of $\Phi_i$ at infinity}

 Let us finish this Appendix by  discussing
the behaviour of $\Phi_i(\w)$ at infinity.  Again we carry out the
argument for $\Phi_1=\Phi_1$ only. The Laplace transform
$\overline\Phi_1 (s,\theta)$ has an asymptotic expansion similar to
(\ref{tipasymp}), with the expansion coefficients that are functions
of $\theta$
 (Fedoryuk 1977 and and Osipov and Norris 1999, Theorem 2).  Significantly,
the $p=0$ terms are either absent or involve constant coefficients
(Kozlov {\it et al.} 2001). Therefore, the function
\begin{eqnarray}
\Phi_1'(\w) = \frac{1}{2} [\partial_\w \Phi_1(\w+\theta)+\partial_\theta
\Phi_1(\w+\theta)]\Big |_{\theta=0}
\end{eqnarray}
has the expansion of the type (\ref{tipasymp}) without a constant
term.
 Using (\ref{change_var}) this implies that as $\Imag \w \rightarrow \infty$, $\Phi_1'(\w) \rightarrow 0$ and therefore,    $\Phi_1(\w) \rightarrow $ const.
When the constant is chosen to be zero (or another specified value)
we call the corresponding function $\Phi_1(\w)$ the Sommerfeld
amplitude.

\section{The tip asymptotics of the elastic potentials and
asymptotics of the Sommerfeld amplitudes at infinity} \label{tip}
The behavior of solutions of the elliptic problems in regions with
piecewise smooth boundaries has been studied by many authors (see
Nazarov and Plamenevskij 1994 and references therein). A rigorous
theory has been developed after a breakthrough by Kondrat'ev (1963)
who constructed and justified the field asymptotics in the vicinity
of conical and edge points. The theory implies that the solution of
the underlying Lam{\' e}   problem
 must have the
asymptotic expansion
\begin{equation} \label{gen_ser}
 \u(kr,\theta)\sim\sum_{\ell,m=0}^\infty (kr)^{q_m+2\ell}\sum_{n=0}^{N_m-1}
\u_{\ell,m,n}(\theta)\, ( \ln kr)^n,\ \
 \ kr\rightarrow  0,
\end{equation}
where for one $m$,  $q_m = 0$ and  $N_m=1$ (otherwise, the tip
conditions are violated); and for any other $m$, $\Real q_m
>0$
and $q_m$ is a root of a transcendental equation,
 with
a natural number $N_m$   being  its multiplicity.

The asymptotic expansions \eqref{gen_ser} can be differentiated and
substituted into the boundary conditions. Therefore, using
\eqref{pot_via_disp} similar expansions may be written for the
elastodynamic potentials $\phi^\pm_i(kr,\theta)$ as
\begin{eqnarray}
\phi^\pm_i (kr, \theta) \sim \sum\limits_{\ell,m=0}^\infty
(kr)^{p^\pm_m+2\ell}\sum\limits_{n=0}^{N_m^\pm-1}
\phi^\pm_{i,\ell,m,n}(\theta) (\ln kr)^n ,\,\,\, kr \rightarrow 0.
\label{tipasymp}
\end{eqnarray}

 Let us arrange  the sets of exponents $\{p^+_m,\ m=0,1,...$\} and
$\{p^- _m,\ m=0,1,...\}$, each in order of the increasing real part.
Following Kozlov {\it et al.}~(2001), these sets may be described as
follows: Each contains $1$, while any other element is a root of the
transcendental equation
\begin{equation}
(p^\pm+1)~{\rm sin~} 2 \alpha \pm {\rm sin~} 2\alpha (p^\pm+1) = 0
 \label{tran_sym}
\end{equation}
and satisfies condition
\begin{equation}
\Re p^\pm >-1 \label{p_m_est}
\end{equation}
(otherwise, the tip conditions are violated). Note that if in
(\ref{gen_ser}),  a $q _m
 \neq 0$, then in (\ref{tipasymp})  the corresponding  $p^+_m$ or
$p^-_m$ equals $q_m-1$, but applying the nabla operator to the term
with $q_m=0$ and $\ell=0$  always gives us zero---because the
corresponding $N_m=1$. Thus, for $q_m=0$, only the next, $r^2$, term
in (\ref{gen_ser}) gives rise to a nonzero term in (\ref{tipasymp}).
The corresponding $p^\pm_m=1$. Note that while for any wedge angle
there exists an $m$, such that $p^-_m=0$ is a solution of
(\ref{tran_sym}), in our range of wedge angles $2\alpha \in (0,
\pi)$, all $p^+_m$ differ from zero. The full set of solutions of
the transcendental equations (\ref{tran_sym}) is described in Kozlov
{\it et al.} (2001). The main facts can be summarized in plots
representing roots of the transcendental equations as functions of
the wedge angle $2\alpha$ (see e.g.~Ting, 1984). The analysis of
these plots shows that in our range of  wedge angles, the tip
conditions are assured for those non-negative exponents with the
minimal real part that are either $1$ or else are  solutions of the
corresponding transcendental equations, with the real part less or
equal than $1$. In other words, all leading exponents in
(\ref{gen_ser}), that is, the exponents with the minimal real part
lie in the strip
\begin{eqnarray}
 0\le {\rm Re}~ p^\pm_0 \le 1. \label{p_int}
\end{eqnarray}
The root loci in Ting (1984) also show that at one wedge angle,
$2\alpha_*\approx0.8\pi$, we have a degeneracy: In the corresponding
symmetric problem, the
 exponent with the minimal real part, $p^+_*\approx 0.76$ is a
multiple root of the corresponding transcendental equation
(\ref{tran_sym}).  The corresponding multiplicity $N_*^+=2.$ There
are no multiple roots $p^-_0$ which have the minimal real part and
simultaneously satisfy (\ref{p_int}).  It follows that for
$\alpha_*$, the leading terms in (\ref{gen_ser}) are
\begin{eqnarray}
\phi^+_{i,0,1,1}(kr)^{p^+_*}{\rm ln}~ kr =O\bigl ((kr)^p \bigr ),
\,\,\,0<p<p^+_*. \label{log_term}
\end{eqnarray}

The behavior of  potentials $\phi^\pm_i(kr,\theta)$  in the vicinity
of the wedge tip dictates the asymptotic behavior of the Sommerfeld
amplitudes $\Phi^\pm_i(\w)$  at infinity: For example, it is easy to
check that for any small $\ve>0$, as $\Imag\w \rightarrow \infty, $
$\Phi^\pm_i(\w)$ have expansions
\begin{eqnarray}
&&\Phi^\pm_i(\w) \sim \sum_{0\le\Re p^\pm_m\leq 1}{
\Phi}^\pm_{im}\re^{ \ri p^\pm_m\w}+O(\re^{\ri(1+\ve)\w}),
\,\,\,\,\alpha \neq \alpha_*,\ 0<\alpha<\frac{\pi}{2}, \nonumber\\
&&\Phi^+_i(\w) \sim {\overline \Phi}_{i*}\w\re^{\ri p^+_*\w}+
{\Phi}_{i*}\re^{ \ri p^+_*\w}+{\Phi}_{i1}\re^{
 \ri\w}+O(\re^{\ri(2+\ve)\w}),\,\,\,
\alpha = \alpha_*. \label{aspp3}
\end{eqnarray}
Note that in the symmetric case, the expansions contain no constant
terms (so that the leading exponents are $1$ and possibly a solution
of \eqref{tran_sym}, with the real part in $(0,1)$), but these may
be present in the antisymmetric case (so that the leading exponents
there are $0$ and $1$).

\section{Radiation conditions at infinity} \label{app:limabs}

In wedge problems, the radiation conditions at infinity are readily
expressed in the form of the limiting absorbtion principle: As
$kr\rightarrow \infty$, the  diffracted waves ${\bf
u}^{\text{diff}}$ with $\Imag k<0$ should decay exponentially. The
solution of the original problem is obtained by taking the limit as
$\Imag k\rightarrow  0$.

Recently a version of the limiting absorbtion principle for the
elastic wedge has been justified in Kamotski and Lebeau (2006). As a
result, the new radiation conditions have been introduced which are
consistent with the physics of the dynamic problem under
consideration. The new radiation conditions have allowed the authors
to prove that the elastic wedge problem stated in terms of
displacements has a unique solution. Reverting to the language of
potentials the Kamotski and Lebeau conditions may be re-written as

\noindent {\bf the inner radiation conditions}
\begin{align}\label{rc_i_1}
\intlim_{|\theta|<\alpha-(kr)^{-1+\varepsilon}}\left|\ri\gamma
k\phi_0^{\text{diff}}+ \partial _r\phi_0^{\text{diff}}
\right|^2\,r\rd\theta\rightarrow 0,\\
\intlim_{|\theta|<\alpha-(kr)^{-1+\varepsilon}}\left|\ri
k\phi_1^{\text{diff}} +\partial_r
\phi_1^{\text{diff}}\right|^2\,r\rd\theta\rightarrow  0,\ &&
kr\rightarrow \infty
\end{align}
{\bf surface radiation conditions}
\begin{align}\label{rc_b}
\int\limits_{\alpha-(kr)^{-1+\varepsilon}\le|\theta|\le\alpha}\left|\ri
k_R \phi_i^{\text{diff}}+\partial_r
\phi_i^{\text{diff}}\right|^2\,r\rd\theta\rightarrow  0, &&
r\rightarrow \infty,
\end{align}
where $k_R=\Omega/c_R$ plus the requirement that as $kr\rightarrow
\infty$ the diffracted field satisfies  the {\bf boundedness
condition}. Following  the methods of Kamotski and Lebeau (2006) it
must be possible to show that the same conditions  are sufficient
for proving uniqueness of the elastic wedge diffraction problem
stated in terms of elastodynamic potentials.

\section{Geometrico-elastodynamic  poles of $\Phi_i$}
\label{app:GEpoles}

In this Appendix we present recursive formulae for evaluating the
geometrico-elastodynamic  poles of $\Phi_i(\w)$ with the
corresponding residues.
\subsection{The symmetric problem}

Given an incident P (compressional) wave
\begin{equation}
\label{inc_sca} \phi_0^{\text{inc}}(kr,\theta)=\re^{\ri[\gamma
kr\cos(\theta-\theta_0^{\text{inc}})]},
\end{equation}
 the symmetric function
$\Phi_0^+(\w)$ has two poles at the angles of incidence $\theta_{01}$ and
$\theta_{02}$, with the corresponding residues $\Res \Phi_0^+(\theta_{0\ell}), \,\ell=1,2$.
These are
\begin{eqnarray}
&&\theta_{01}=-\theta_0^{\text{inc}},~~\theta_{02}=\theta_0^{\text{inc}}, \nn \\
&&\Res \Phi_0^+ (\theta_{0\ell}) =  -\frac{1}{4 \pi \ri}. \label{polef}
\end{eqnarray}
 The poles lie in the strip $|\Re \w|\le\alpha$. It follows
from the first equation in \eqref{san_cont} that $\Phi_0^+(\w)$ also
has P-P poles $\theta_{0\ell},\,\ell=3,4$, with the respective residues $\Res
\Phi_0^+(\theta_{0\ell})$.  These describe the once reflected compressional
wave  and are given by
\begin{eqnarray}
&&\theta_{0\ell} =\theta_{0,\ell-2}+ 2\alpha,\nn\\
&&\Res \Phi_0^+(\theta_{0\ell}) = {\rm
r}_{11}(g^{-1}(\theta_{0,\ell-2}+\alpha))\Res \Phi_0^+ (\theta_{0,\ell-2}).
\label{pol1} \end{eqnarray} The S (shear) waves generated on the
first reflection are associated with the P-S poles $\w =
\theta_{1\ell},\,\ell=3,4$, with the corresponding residues $\Res
\Phi_1^+(\theta_{1\ell}).$ These are given by
\begin{eqnarray}
&&\theta_{1\ell} =
g^{-1}(\theta_{0,\ell-2}+\alpha)+\alpha\nn\\
&&\Res \Phi_1^+(\theta_{1\ell}) = \frac{{\rm
r}_{21}(g^{-1}(\theta_{0,\ell-2}+\alpha))}{g'(\theta_{1\ell}-\alpha)}\Res
\Phi_0^+ (\theta_{0,\ell-2}). \label{pole2}
\end{eqnarray}
Similarly,  the poles $\theta_{0\ell}$ and $\theta_{1\ell},\,\ell=5,6$  of the type
P-P-P and P-P-S are described by formulae (\ref{pol1}) and
(\ref{pole2})  respectively. The   P-S-P poles $\theta_{0\ell},\,\ell=7,8$ are
given by
\begin{eqnarray}
&&\theta_{0\ell} = g(\theta_{1,\ell-4}+\alpha)+\alpha\nn\\
&& \Res \Phi_0^+(\theta_{0\ell}) = {\rm
r}_{12}(\theta_{1,\ell-4}+\alpha)g'(\theta_{1,\ell-4}+\alpha)\Res \Phi_1
(\theta_{1,\ell-4}),
\end{eqnarray}
and P-S-S poles  $\theta_{1\ell},\,\ell=7,8$ by
\begin{eqnarray}
&&\theta_{1\ell}
=\theta_{1,\ell-4}+ 2\alpha,\nn\\
&&\Res \Phi_1^+(\theta_{1\ell}) = {\rm r}_{22}(\theta_{1,\ell-4}+\alpha)\Res
\Phi_1^+ (\theta_{1,\ell-4}). \label{polel}
\end{eqnarray}

The above formulae should be applied recursively until one evaluates
all poles within  (\ref{A}). Due to the symmetry of $\Phi_0^+(\w)$
and $\Phi_1^+(\w),$ $-\theta^{ik}$ are their respective poles as
well, with the corresponding residues ${\rm Res~} \Phi_0^+ (-
\theta_{0\ell}) = {\rm Res~} \Phi_0^+ ( \theta_{0\ell})$ and ${\rm Res~}
\Phi_1^+ (- \theta_{1\ell}) = -{\rm Res~} \Phi_1^+ ( \theta_{1\ell})$. If the
recursive procedure is initiated outside the strip  $|\Re
\w|\le\alpha$, then, as equations (\ref{san_cont}) suggest, other
pairs of poles appear to be possible. However,  direct calculations
show that such points give a zero contribution.

\subsection{The antisymmetric problem}

The antisymmetric problem allows a similar treatment, the analytic
continuation should also be performed using  Eq. (\ref{san_cont})
with the ``$-$'' superscipt.

\section
{The integral equations for one unknown}\label{inteqns}

\subsubsection*{\it Symmetric problem}

 The two final regularized integral equations to solve are
\begin{eqnarray}
\label{app_int_eq_main^+} \widetilde y^+_i(\eta)+\widetilde
L^+\widetilde y^+_i(\eta)=\widetilde q^+_i(\eta), \,\,i=0,1,
\end{eqnarray}
where $\widetilde L^+$ is an operator with a smooth kernel
\begin{eqnarray} &(\widetilde
L^+u)(\eta)&=-\frac{1}{4\alpha^2}\intlimR{l^+(\eta,t)\tanh
  2t\over\sqrt{d(t)}\sqrt{d(\eta)}\cosh
  \palpha\eta}u(t)
\,dt,
\end{eqnarray}
with
\begin{eqnarray} l^+(\eta,t)= \text{V.P.}\intlimR{\cosh\palpha \tau\over
  \sinh\palpha(\tau-\eta)}
 \Big\{{\tanh
  2t\over\chi'(t)\sinh\palpha  (t-\tau)}-
{\tanh 2\tau\over\sinh\palpha
[\chi(t)-\chi(\tau)]}\Big\}\,d\tau.\label{l+} \nonumber
\end{eqnarray}
The respective right hand sides of Eqs. \eqref{app_int_eq_main^+}
are given by
\begin{eqnarray}
\widetilde q^+_i(\eta)=\frac{1}{\displaystyle 2\alpha
\ri\sqrt{d(\eta)} \cosh\palpha \eta}
\text{V.P.}\displaystyle\intlimR\frac{\cosh\palpha  t}
{\sinh\palpha(t-\eta)}q^+_i(t)\,dt,\,\, \label{rhss+}
\end{eqnarray}
with $q_i^+(\eta)$ given   in Appendix \ref{nomenclature}. On
solving  \eqref{app_int_eq_main^+},   $y^+(\eta)$ is obtained using
\begin{equation}
y^+(\eta)=d^{-1/2}(\eta)[\widetilde y^+_0(\eta)+c^+_1\widetilde
y^+_1(\eta)],
\end{equation}
where $c^+_1=-\lambda^+_0/\lambda^+_1$ and we have
\begin{eqnarray}
&&\lambda^+_i=\intlimR\bigl[\widetilde y^+_i(t)d^{-1/2}(t)
(B^+{\mathbb I})(t)-q^+_i(t)\bigr]dt,\,\,\,i=0,1,\nonumber\\
&& (B^+{\mathbb I})(t)=\frac{\gamma^2\tanh 2t}{2\alpha \ri}\intlimR
\frac{\sinh
2(\chi(t)+\tau)}{1+2\gamma^2\sinh^2(\tau+\chi(t))}\,\cdot\,
  \frac{d\tau}{\sinh \palpha \tau}.\nonumber
\end{eqnarray}

\subsubsection*{\it Antisymmetric problem }

Analogously to  the symmetric case, the two final regularized
integral equations to solve are
\begin{eqnarray}
\label{app_int_eq_main^-} \widetilde x_i^-(\eta)+\widetilde
L^-\widetilde x_i^-(\eta)=\widetilde q_i^-(\eta),\,\,i=0,1,
\end{eqnarray}
where the integral operator is
\begin{eqnarray}
\ \ \ &(\widetilde L^-u)(\eta)&=-
\frac{1}{4\alpha^2}\intlimR{l^-(\eta,t)\tanh
  2t\over\sqrt{d(t)}\sqrt{d(\eta)}\cosh
  \palpha\chi(\eta)}u(t)
\,dt,
\end{eqnarray}
with
\begin{eqnarray}
\label{l-} \ \ \ l^- (\eta,t)= -\text{V.P.}\intlimR{\cosh\palpha
\tau\over
  \sinh\palpha[\chi(\eta)-\tau]}
\Big\{{\tanh 2t\over\sinh\palpha (\chi(t)-\tau)}+
\frac{\tanh2\chi^{-1}(\tau)({{\chi^{-1}}})'(\tau)}
{\sinh\palpha[\chi^{-1}(\tau)-t]}\Big\}\,d\tau.\nonumber
\end{eqnarray}
The respective right hand sides of Eqs. \eqref{app_int_eq_main^-}
are given by
\begin{eqnarray}
\widetilde q_i^-(\eta)=\frac{1}{2\displaystyle \alpha
\ri\sqrt{d(\eta)}\cosh\palpha\chi(\eta)} \text{V.P.}\displaystyle
\intlimR\frac{ \cosh\palpha t}
{\sinh\palpha[t-\chi(\eta)]}q_i^-(\chi^{-1}(t))\,dt,\,\,
\label{rhss-}
\end{eqnarray}
with $q_i^-(\eta)$  given   in Appendix \ref{nomenclature}. As
above, $x^-$ is obtained on solving \eqref{app_int_eq_main^-} using
\begin{eqnarray}
x^-(\eta)=d^{-1/2}(\eta)[\widetilde x_0^-(\eta)+c_1^-\widetilde
x_1^-(\eta)],
\end{eqnarray}
where $c_1^-=-\lambda_0^-/\lambda_1^-$ and we have
\begin{eqnarray}
&&\lambda_i^-=\intlimR \bigl [ \widetilde x_i^-(t)d^{-1/2}(t)
(B^-\chi') (t)-q_i^-(t)\chi'(t) \bigr]dt,
\,\,i=0,1,\nonumber\\
&& (B^-\chi')(t)=\frac{\tanh 2t}{2\alpha \ri}\int_{- \infty}^\infty
\frac{\tanh 2(t+\tau)\, d\tau}{\sinh \palpha \tau}.\nonumber
\end{eqnarray}

\section{The  branch points of $\Phi_i(\w)$} \label{app:Rploes_brcuts}

Let us show that the Sommerfeld amplitudes $\Phi^+_i(\w)$ that
satisfy the functional equations (\ref{fe1}) and
conditions(\ref{sas_deco}) can have only physically meaningful
branch points. For simplicity of presentation, let us assume that
$\theta_h<2\alpha$ (in the opposite case, a slightly more involved
argument still goes through.) Using the branch points of $g(\w)$,
the only branch points that  $\Phi_1^+(\w)$ can have inside
(\ref{A})
 are $\pm(\pi +
\alpha - \theta_h)$. The corresponding branch cuts run along the
segments $[-\pi -\alpha - \theta_h,~ -\pi - \alpha + \theta_h]$ and
$[\pi + \alpha - \theta_h, ~\pi + \alpha + \theta_h]$.  Note that
the branch points $\pm(\pi+\alpha+\theta_h)$  lie outside the
physical region (\ref{A}).  The analogous branch points of
$\Phi_0^+(\w )$ can be only $\pm(\pi + \alpha \pm \ri {\cosh^{-1}}
(1/\theta_h))$, with the cuts along the segments $[-\pi -
\alpha+\ri{\cosh^{-1}~} (1/\theta_h) ,~ -\pi - \alpha
-\ri{\cosh^{-1}~} (1/\theta_h)]$ and $[\pi + \alpha
+\ri{\cosh^{-1}~} (1/\theta_h), ~\pi + \alpha - \ri{\cosh^{-1}~}
(1/\theta_h)]$.

Indeed, all possible branch points of $g(\w)$ are $\pi n \pm
\theta_h$ (see \eqref{gcuts}). In principle, applying (\ref{fe1}),
they could generate many branch points in $\Phi_1^+(\w)$ which have
no physical interpretation. Let us start by showing that
$-\alpha+\theta_h$
 is not a branch point: Let us use the fact that $\Phi_0^+(\w)$
is odd  to rewrite  the functional equation (\ref{fe1}) as
\begin{equation}
t_{11}(\w+\alpha)\{\Phi_0^+[\alpha+g(\w+\alpha)]-\Phi_0^+[\alpha-
g(\w+\alpha)]\}+t_{12}(\w+\alpha)[\Phi_1^+(\w+2\alpha)+\Phi_1^+(\w)
]= Q_1^+, \label{fe0_2}
\end{equation}
\begin{equation}
t_{21}(\w+\alpha)\{\Phi_0^+[\alpha+g(\w+\alpha)]+\Phi_1^+[\alpha-
g(\w+\alpha)]\}+t_{22}(\w+\alpha)[\Phi_1^+(\w+2\alpha)-\Phi_1^+(\w)
]= Q_1^+. \label{fe1_2}
\end{equation}
In the vicinity of $\theta_h$,  there exist the constants $a_n$ such
that we have
\begin{equation}
g(\w+\alpha)=\sum_{n=0}^{\infty} a_n (\w+\alpha-\theta_h)^{n+1/2}.
\end{equation}
Since $\Phi_0^+[\alpha+g(\w+\alpha)]-\Phi_0^+[\alpha-g(\w+\alpha)]$
is odd in $g$, there also exist constants $A_n$ such that that this
function has the expansion
\begin{equation}
\Phi_0^+[\alpha+g(\w+\alpha)]-\Phi_0^+[\alpha-g(\w+\alpha)]=\sum_{n=0}^{\infty}
A_n (\w+\alpha-\theta_h)^{n+1/2}.
\end{equation}
On the other hand, there exist constants $b_n$ such that we can
write
\begin{equation}
t_{11}(\w+\alpha)=\sum_{n=0}^{\infty} b_n
(\w+\alpha-\theta_h)^{n-1/2}.
\end{equation}
Thus, the first term in the left hand side  of (\ref{fe0_2})
contains no branch points. The coefficient $t_{12}$ has no branch
points either.  It follows that there are no branch points in
Eq.~(\ref{fe0_2}) at all.

Let us move on to Eq.~(\ref{fe1_2}). The function
$\Phi_0^+[\alpha+g(\w+\alpha)]+\Phi_0^+[\alpha-g(\w+\alpha)]$ is an
even function of $g$, and therefore, there exist constants $B_n$
such that we can write
\begin{equation}
\Phi_0^+[\alpha+g(\w+\alpha)]+\Phi_0^+[\alpha-g(\w+\alpha)]=\sum_{n=0}^{\infty}
B_n (\w+\alpha-\theta_h)^n.
\end{equation}
Since the coefficients $t_{21}(\w)$ and $t_{22}(\w)$ have no branch
points, there are no branch points in Eq.~(\ref{fe1_2}). It follows
that the point $-\alpha+\theta_h$ which does not have any physical
interpretation is not a branch point of the function $\Phi_1^+(\w)$.
Similarly, it can be shown that the points $-\alpha-\theta_h$ and
$\alpha\pm\theta_h$ are not branch points of $\Phi_1^+(\w)$.
Analogous considerations apply in the antisymmetric case.

We conclude that the branch points of the Sommerfeld amplitudes of
the solution of the original problem that lie in the physical region
(\ref{A}) and therefore give rise to physical waves, lie outside the
strip $\Re \w\,\epsilon\,I$.  This means that they do not have to be
taken into account in the functional equations  for
$\tPhi^\pm_i(\w)$  or by the same token, in the resulting singular
integral problem.  On the other hand, we have no theoretical proof
that  $\Phi^\pm_i(\w)$ that we eventually compute have only physical
branch points in the physical region(\ref{A}). We can confirm this
fact only by carrying out numerical tests.

\section{The Rayleigh
reflection and transmission coefficients}\label{coefs}

When evaluating (\ref{S2int3}), a pole $\theta^{{\rm inc}}_{1{\rm
R}}=\alpha-\ri \beta_{\rm R}$ of $\Phi_1(\w)$, with $\beta_{\rm
R}>0$ corresponds to a plane wave with the phase factor
\begin{eqnarray}
\re^{\ri kr \cos (\theta-\theta^{{\rm inc}}_{1{\rm R}})}= \re^{\ri
kr \cos (\theta-\alpha )\cosh \beta_{\rm
R}}\re^{-kr{\sin(\alpha-\theta) \sinh \beta_{\rm R}}},
\end{eqnarray}
so that its amplitude is exponentially small everywhere except for a
small neighborhood of the wedge face $\theta = \alpha$. Thus, we
describe a  Rayleigh  wave incident from infinity  along the upper
 face $\theta = \alpha$  by two potentials
\begin{eqnarray}
\phi^{\text{inc}}_i(kr,\theta)=4\pi \ri\phi_{i0}\re^{\ri\gamma
kr{\rm cos}(\theta -\theta^{{\rm inc}}_{i{\rm R}})},
\end{eqnarray}
 where $
\phi_{00}=1$ and $\phi_{10}={\displaystyle -2\ri\gamma_{\rm
R}\sqrt{\gamma_{\rm R}^2- \gamma^2}}/({\displaystyle 2\gamma_{\rm
R}^2-1})$; $\gamma_{\rm R} = c_{\rm S}/c_{\rm R}$ with $c_{\rm S}$
being the Rayleigh wave speed  and $\theta^{{\rm inc}}_{0{\rm R}}=
\alpha-g(\ri\beta_{\rm R})$. The reflected wave propagates along the
same wedge face as the incident but from the tip to infinity and
transmitted---along the other face, again away from the tip. They
are described respectively by
\begin{eqnarray}
&&\phi^{\rm ref}_i(kr,\theta)=4\pi \ri R^{\rm ref}
\phi_{i0}\re^{-\ri\gamma kr{\rm cos}(\theta-\theta^{{\rm sc}}_{i{\rm
R}}
)},\nonumber\\
&&\phi^{\rm tran}_i(kr,\theta)=4\pi\ri R^{\rm tran}
\phi_{i0}\re^{-\ri\gamma kr{\rm cos}(\theta+\theta^{{\rm sc}}_{i{\rm
R}})},
\end{eqnarray}
with ``scattering angles'' $\theta^{ {\rm sc}}_{i{\rm R}}$ being the
complex conjugate of $\theta^{{\rm inc}}_{i{\rm R}}$, so that
$\theta^{{\rm sc}}_{0{\rm R}}= \alpha+ g(\ri\beta_{\rm R})$ and
$\theta^{{\rm sc}}_{1{\rm R}}= \alpha+ \ri\beta_{\rm R}$. Above,
$R^{\rm ref}$ and $R^{\rm tran}$ are the Rayleigh reflection and
transmission coefficients
\begin{eqnarray}
R^{\text{\rm
ref}}=\frac{1}{2}[R^{\text{+ref}}+R^{-\text{ref}}],\,\,\,
R^{\text{\rm tran}}=\frac{1}{2}[R^{\text{+ref}}-R^{-\text{ref}}],
\end{eqnarray}
with the symmetric and antisymmetric  parts given respectively by
\begin{eqnarray}
R^{\pm\text{\rm ref}}=\text{Res}[\Phi_0^\pm;\,g(\w_{\rm
R})+\alpha],\,\,\,{\rm with~}\w_{\rm R}=\pi+\ri\beta_{\rm R},
\end{eqnarray}
so that using the additional angles $\theta_{0,{\rm
R}}=-\alpha+g(\w_{\rm R})$ and $\theta_{1,{\rm R}}=-\alpha+\w_{\rm
R}$, we have
\begin{eqnarray}
&&R^{\pm\text{ref}} = \pm\sum_{k=1}^2 {\rm r}_{1k}(\w_{\rm
R})\Phi^\pm_{k-1}(\theta_{k-1,{\rm R}}) + c_1^\pm e^\pm_{1}(\w_{\rm
R}) \frac{g'(\w_{\rm R})\Delta(\w_{\rm R})}{\Delta'(\w_{\rm
R})}\sqrt{\gamma^2-{\rm cos}^2\w_{\rm R}}.\nonumber
\end{eqnarray}
As before,  the dash denotes the derivative with respect to the
argument.
\section{Adjustable functions and parameters in singular terms}\label{numerics}

 As any other
numerical code ours relies on a choice of certain options which
effect a tradeoff between numerical accuracy and either running time
or else numerical stability.  Apart from the relevant grids, these
options are

\begin{remunerate}
\item[(i)] {\it The adjustable function
$\sigma(\w)$ in (\ref{*dec})} In most cases, $\sigma(\w)$ is chosen
to be
\begin{equation}
\sigma(\w)
 = \frac{\frac{\displaystyle \pi}{\displaystyle 2\alpha}}
{\sin\frac{\displaystyle \pi}{\displaystyle 2\alpha}\w}
\label{perfect}
\end{equation}
({\it cf.} Budaev and Bogy 1995, Eq.~(15)). The choice is
convenient, because it simplifies the right hand sides of our
functional  and therefore,  integral equations. Also, $\sigma(\w) $
in (\ref{perfect}) decays at infinity reasonably fast. However, for
any wedge angle $2\alpha$, there exists a critical incident angle
$\theta^{\text{inc}}_0$, such that one of the poles of
$\Phi^\pm_i(\w)$ lies on the boundary of the strip $\Re
\w\,\epsilon\,I$. For illustration purposes, let it be
$\widehat\Phi^\pm_1(\w)$ and let the pole be $\theta_0 =
\pi/2-\alpha$. Then the corresponding term ${\rm Res~}(
\Phi^\pm_1;\theta_0) \sigma(\w-\theta_0)$ has one more nonphysical
pole, $\theta_0 + 2\alpha $. In situations like these, another
choice of $\sigma(\w)$ is called for, with  poles further apart.  We
have tested
\begin{equation}
\sigma(\w)
 = \frac{\beta}
{\sin\beta\w}. \label{pmf1}
\end{equation}
with various values  $\beta <\pi/ (2\alpha)$. However, any
$\sigma(\w)$ different from (\ref{perfect}) leads to more cumbersome
right hand sides of the integral equations and exhibits a slower
decay. As a result, the function (\ref{pmf1}) while increasing the
stability of the solution, increases the code run time roughly
tenfold. For this reason, we abandon \eqref{perfect} only when
$\theta^{\text{inc}}$ is near critical angle. In this region  we use
\eqref{pmf1},  with $\beta=\pi/(6\alpha)$.
 \item[(ii)] {\it Number of
poles.} When evaluating the poles of the Sommerfeld amplitudes we do
not have to restrict ourselves to the strip $\Re \w\,\epsilon\,I$.
The more poles that are utilized in evaluation, the wider the domain
of analyticity of the corresponding unknowns $\tPhi_i(\w) $
  and therefore, the higher the accuracy. On the other
hand, some nonphysical poles possess residues with large amplitudes
and cause numerical instability. In the present version of the code,
when $\theta$ is away from critical we take into account all poles
in the strip $\Re \w\,\epsilon\,I$,   and when  $\theta$ is near
critical we take into account all poles in the wider strip $\Re
\w\,\eps\,[\pi/2-2\alpha,\,\pi/2+2\alpha]$.
\end{remunerate}

\section*{ Acknowledgements}
We
 are grateful to Drs R.K
Chapman, D. Gridin and J. Hudson for many useful and insightful
comments and Professors V. P. Smyshlyaev and A. Gautesen for
fruitful discussions and suggestions. We would also like to thank
Profs. Nazarov and Plamenevskij for invaluable advice and
 relevant references.
\vfill

\newpage

\section{Nomenclature}\label{nomenclature}
\begin{eqnarray}
&&a(\eta)=-\ri\tanh 2\eta, \nonumber\\
&&b(\eta)=\frac{2\ri\sinh\eta\sqrt{\gamma^2+\sinh^2\eta}}{\cosh
2\eta},\nonumber\\
&&d(\eta)=1-\frac{\tanh^22\eta}{\chi'(\eta)},\nonumber\\
&&e^+_{1}(\w) =-\tan\alpha\, \sin 2\w +\cos2\w,\nonumber\\
&&e^-_{1}(\w)=\tan\alpha\,\cos2\w  +\sin2\w,\nonumber\\
&&e^+_{2}(\w) =\tan\alpha\, \cos 2\w
\frac{\sin \w}{\sqrt{\gamma^2-{\rm cos}^2\,\w}}+\sin2\w,\nonumber\\
&&e^-_{2}(\w)=\tan\alpha\,\sin 2\w -
\cos 2\w \frac{\sin \w}{\sqrt{\gamma^2-{\rm cos}^2\,\w}},\nonumber\\
&&g(\w)= {\rm cos}^{-1}(\gamma^{-1}\cos \w),\nonumber\\
&&{\mathcal H}f(\eta)= \frac{1}{2\alpha \ri} \text{V.P.}\intlimR
\frac{f(t)\rd
t}{\sinh[\frac{\pi}{2\alpha}(t-\eta)]}, \nonumber\\
&&({\mathcal H}^{-1}f)(\eta)=\frac{1}{2\alpha
\ri}\text{V.P.}\intlimR
\coth\palpha(t-\eta)f(t)\,dt,\nonumber\\
&&{\overline {\mathcal H}} f(\eta)= \frac{1}{2\alpha \ri}
\text{V.P.}\intlimR \frac{f(t)\chi'(t)\rd
t}{\sinh\palpha[\chi(t)-\chi(\eta)]},
\nonumber\\
&&Kf(\eta)=\frac{1}{2\alpha \ri}\intlimR
\left\{\frac{\tanh^22t}{\chi'(t)\sinh\palpha(t-\eta)}- \frac{\tanh
2t\tanh 2\eta} {\sinh\palpha[\chi(t)-\chi(\eta)]}\right\}
f(t)\,dt,\nonumber\\
&&q_0^+(\eta)=r_2^+(\eta)-a(\eta){\overline {\mathcal H}}r_1^+(\eta),\nonumber\\
&&q_1^+(\eta)= - \tan \alpha\frac{\cosh \eta}{\cosh
2\eta}-\frac{\tanh2\eta}{2\alpha} \text{V.P.}\intlimR
\frac{\gamma\cosh\tau d\tau}{(1+2\gamma^2\sinh^2\tau)
\sinh\palpha[\tau-\chi(\eta)]}, \nonumber\\
&&q_0^-(\eta)=r_1^-(\eta)-b(\eta){\mathcal H}r_2^-(\eta),\nonumber\\
&&q_1^-(\eta)= - \tan \alpha\frac{\cosh \eta}{\chi'(\eta)\cosh
2\eta} -\frac{\tanh 2\eta}{2\alpha\chi'(\eta)} \text{V.P.}\intlimR
 \frac{\cosh\tau\, d\tau}{\cosh 2\tau\,\sinh\palpha(\tau-\eta)},  \nonumber\\
&&r_1^\pm(\eta) =- \Big[ \widehat\Phi_0^+
(g(\frac{\pi}{2}+\ri\eta)+\alpha)\pm \widehat\Phi_0^+
(g(\frac{\pi}{2}+\ri \eta)- \alpha)\Big]- b(\eta) \Big[
\widehat\Phi_1^+ (\frac{\pi}{2}+\alpha+\ri\eta) \pm \widehat\Phi_1^+
(\frac{\pi}{2}-\alpha+\ri\eta)
\Big],  \nonumber\\
&&r_2^\pm(\eta) =- a(\eta)\Big[ \widehat\Phi_0^+
(g(\frac{\pi}{2}+\ri\eta )+\alpha)\mp \widehat\Phi_0^+
(g(\frac{\pi}{2}+\ri\eta )-\alpha)\Big]- \Big[ \widehat\Phi_1^+
(\frac{\pi}{2}+\alpha+\ri\eta)\mp\widehat\Phi_1^+(\frac{\pi}{2} - \alpha+\ri\eta)\Big],\nonumber\\
&&{\rm r}_{11}(\w)=-{\rm r}_{22}(\w)=\frac{2\sin 2\w~\cos
\w\sqrt{\gamma^2-{\rm
cos}^2~\w}-{\rm cos}^2~ 2\w}{\Delta(\w)},\nonumber\\
&&{\rm r}_{12}(\w)=-\frac{4\cos 2\w~\cos \w\sqrt{\gamma^2-
{\rm cos}^2~\w}}{\Delta(\w)}, ~~\nonumber\\
&&{\rm r}_{21}(\w)= -\frac{2\sin 2\w~\cos 2\w}{\Delta(\w)},\nonumber\\
&&\Delta(\w)={\rm cos}^2 ~2\w+2~\sin 2\w~\cos \w\sqrt{\gamma^2-{\rm
cos}^2~\w},\nonumber\\
&&\chi(\eta)=\sinh^{-1}(\gamma^{-1}\sinh(\eta)),\nonumber\\
&&\chi'(\eta)=\frac{\cosh\eta}{\sqrt{\gamma^2+\sinh^2\eta}}.
\nonumber
\end{eqnarray}
Note that the Rayleigh function $\Delta(\omega)$ has the purely
imaginary root $\ri\beta_{\rm R}$, with $\beta_{\rm R}>0$.

\end{document}